\documentclass{aa}
\usepackage{graphicx, graphics}
\usepackage{CJK}
\usepackage{bbm}
\usepackage{amsmath, amssymb, bm}
\usepackage[utf8]{inputenc}
\usepackage[normalem]{ulem}
\usepackage[nodayofweek]{datetime}

\input{prepcitationformat.txt}
\usepackage{txfonts}
\usepackage{colortbl}
\usepackage[utf8]{inputenc}

\usepackage{pdflscape}
\usepackage{rotating}

\usepackage[switch]{lineno}
\usepackage{txfonts}
\usepackage{booktabs}
\usepackage[normalem]{ulem}
\usepackage{multirow}
\usepackage{hyperref}
\usepackage{orcidlink}
\usepackage{graphicx}
\hypersetup{
    colorlinks = true,
    allcolors = {blue}
}
\usepackage{txfonts}
\usepackage{tikz}

\definecolor{lime}{HTML}{A6CE39}
\DeclareRobustCommand{\orcidicon}{%
    \raisebox{-3pt}{\begin{tikzpicture}
    \filldraw [lime, yshift=-2pt] (0, 0) circle [radius=0.16]
    node[white] {\raisebox{1pt}{\hspace{0.5pt}\fontfamily{qag}\selectfont\tiny i\scalebox{0.8}{D}}};
    \end{tikzpicture}}
    \hspace{-2.5mm}
    \vspace{-0.25pt}
}

\newdateformat{monthyearday}{%
  \THEYEAR\ \monthname[\THEMONTH] \twodigit{\THEDAY}}
\newdateformat{daymonthyear}{%
  \twodigit{\THEDAY}\ \monthname[\THEMONTH] \THEYEAR}

\newcommand{\orcidauthor}[2]{#2\href{http://orcid.org/#1}{\orcidicon}}
\titlerunning{LkCa~15 protoplanetary disk}
\authorrunning{Swastik et al.}
\begin{document} 
\begin{CJK*}{UTF8}{gbsn}

   \title{Imaging the LkCa~15 system in polarimetry and total intensity without self-subtraction artefacts\thanks{The final data products in fits format are publicly available at the CDS.} \thanks{Based on observations performed with VLT/SPHERE under program ID 106.21HJ }}


\author{ 
\orcidauthor{0000-0003-1371-8890}{C. Swastik }\inst{\ref{unimi},\ref{inst-eso},\ref{iia},\ref{PU} }\and 
\orcidauthor{0000-0001-8269-324X}{Zahed Wahhaj\inst{\ref{inst-eso}}}  \and 
\orcidauthor{0000-0002-7695-7605}{Myriam Benisty}\inst{\ref{inst-oca}, \ref{inst-uga}} \and
\orcidauthor{}{Saksham Arora\inst{\ref{inst-eso}, \ref{inst-ger}}} \and
\orcidauthor{0000-0002-4438-1971}{Christian Ginski}\inst{\ref{inst-galway}} \and
\orcidauthor{0000-0003-1698-9696}{Bin B. Ren (任彬)} \inst{\ref{inst-oca},\ref{inst-uga}} \and 
\orcidauthor{}{R. G. van Holstein} \inst{\ref{inst-eso}}  
\and Rob de Rosa \inst{\ref{inst-eso}}
\and \orcidauthor{0000-0003-0799-969X}{Ravinder K Banyal }\inst{\ref{iia}} \and \orcidauthor{0000-0003-1451-6836}{Ryo Tazaki
}\inst{\ref{inst-tok}}           
}
\institute{
Dipartimento di Fisica, Universit\'a degli Studi di Milano, Via Celoria 16, 20133 Milano, Italy ; \url{swastik.chowbay@unimi.it} \label{unimi} \and 
European Southern Observatory, Alonso de C\'ordova 3107, Vitacura Casilla 19001, Santiago, Chile \label{inst-eso} \and
Indian Institute of Astrophysics, Koramangala 2nd Block, Bangalore 560034, India ; \label{iia} 
 \and
Pondicherry University, R.V. Nagar, Kalapet 605014, Puducherry, India \label{PU} \and
Universit\'{e} C\^{o}te d'Azur, Observatoire de la C\^{o}te d'Azur, CNRS, Laboratoire Lagrange, Bd de l'Observatoire, CS 34229, 06304 Nice cedex 4, France; \label{inst-oca}
\and
Universit\'{e} Grenoble Alpes, CNRS, Institut de Plan\'{e}tologie et d'Astrophysique (IPAG), F-38000 Grenoble, France \label{inst-uga}
\and
University of Potsdam, Am Neuen Palais 10, 14469 Potsdam, Germany \label{inst-ger}
\and
School of Natural Sciences, University of Galway, University Road, H91 TK33 Galway, Ireland \label{inst-galway} 
\and 
Department of Earth Science and Astronomy, The University of Tokyo, Tokyo 153-8902, Japan \label{inst-tok}
}


   \date{Received April 15, 2022; accepted \today}
\abstract
{Studying young protoplanetary disks is essential for understanding planet formation, but traditional angular differential imaging introduces self-subtraction artefacts that make their small-scale structure difficult to interpret. We present high-resolution total- and polarized-intensity K$_s$-band images of the LkCa~15 system that are free of such artefacts.}
{LkCa~15 is a young protoplanetary system with a $\sim$160 au disk and previous claims of two protoplanet candidates at 15 and 18 au. We aim to analyse the LkCa~15 protoplanetary disk using high-contrast imaging to search for super-Jupiter planets beyond 20 au and to characterise the dust distribution and grain composition.}
{We used near-simultaneous reference-star differential imaging (RDI, ``star-hopping'') to obtain self-subtraction-free K$_s$-band images beyond 0.1\arcsec. We first modelled the K$_s$-band total- and polarized-intensity images together with ALMA submillimetre continuum maps using \texttt{RADMC-3D} and a two grain-size (micron and mm) compact olivine model. Residual mismatches in the near-IR then motivated us to extract the scattering phase function $S(\theta)$ and polarized fraction $P(\theta)$ from the SPHERE data and compare them with aggregate-scattering models, which pointed to porous CAHP grains in the surface layer and led us to recompute the NIR scattered-light models with CAHP.}
{Our initial two grain-size (micron and mm) olivine model roughly reproduces the observed NIR and ALMA disk morphology, with a flared micron surface layer from $\sim$25-85 au ($H/R\sim0.08$ at 50 au; surface gap $\sim$35-40 au) and a millimetre midplane ring from $\sim$ 55-130 au with a gap at $\sim$ 75-100 au, for $i\sim50^\circ$ and $\mathrm{PA}\sim61^\circ$. The near-IR data, however, are less forward-scattering than the model. From the phase functions we find that $S(\theta)$ rises by $\sim5\times$ from $\theta\sim90^\circ$ to $\theta\sim35^\circ$, while $P(\theta)$ shows a broad sub-Rayleigh peak with $P_{\max}\sim0.35$ near $\theta\sim90^\circ$. These analysis disfavour compact olivine Mie spheres and are better matched by porous aggregates (CAHP-128-100 nm), so we recomputed the NIR scattered-light models with CAHP-128-100 nm grains in the surface layer (retaining compact mm grains for the ALMA continuum), which improves the match to the K$_s$-band morphology and polarization. From the number ratio between the 12 $\mu$m and 2 mm grains we infer a size-distribution slope of $\zeta\sim-2.3$. Although no new candidate planets were detected, we estimate upper mass limits: beyond 200 au, planets more massive than $\sim$1.5 M$_J$ are unlikely, while in the inner disk planets up to $\sim$3.6 M$_J$ could remain undetected.}
{The star-hopping RDI data, together with phase-function diagnostics and \texttt{RADMC-3D} modelling with compact olivine and porous CAHP grains, allow us to reproduce the main observed features of the LkCa~15 system. The number ratio between the 12\,µm and 2\,mm olivine grains further shows that micron-sized grains are under-abundant relative to size distributions in the ISM or debris disks, providing new insight into grain growth and dust dynamics in gas-rich protoplanetary disks.}


   \maketitle
%

\section{Introduction}
Protoplanetary disks, the dense circumstellar gas and dust of $\sim$ 1 to 100 M$_{J}$ encircling young stars are widely recognized as the birthplaces of planetary systems \citep{2017A&A...605A..69T,2019A&A...624A..93B}. The direct imaging of these disks and sometimes planets within them offers an unparalleled window into the dynamic processes of planets in formation. So far, in scattered light, we have imaged only a few disks which have a planet embedded in them, such as PDS~70 which is host of two such protoplanets \citep{2018A&A...617A..44K,2018A&A...617L...2M,2019NatAs...3..749H,2021AJ....161..148W} and AB Aur which may also host a planet embedded in the disk \citep{2022NatAs...6..751C}. We have detected a circumplanetary disk around PDS70~c using ALMA observations \citep{2021ApJ...916L...2B}. Further, observations of several proto-planetary disk systems, such as TW Hya \citep{2016ApJ...820L..40A,2017ApJ...837..132V}, HD 97048 \citep{2016A&A...595A.112G,2017A&A...597A..32V}, and HD 142527 \citep{2012ApJ...754L..31C,2017AJ....154...33A}, have revealed intricate patterns of dust rings. These structures have been captured both in scattered light, using Adaptive Optics (AO) systems like GPI, SCExAO and SPHERE, and in the sub-millimeter range, using interferometers such as ALMA. Hydrodynamical simulations, when integrated with sophisticated radiative transfer models, suggest that planets ranging from sub-Jovian to Jovian mass are capable of producing such substructures within their disks \citep[e.g.][]{2016ApJ...823L...8M,2018MNRAS.477.1270P,2015ApJ...812L..32D,2015MNRAS.453L..73D, 2016ApJ...816...25P,2017ApJ...850..201B}. Consequently, the analysis of disk features like gaps, rings, and spirals not only sheds light on the mass and properties of the associated proto-planets but also offers insights into the evolutionary mechanisms at play. In this context, the LkCa~15 protoplanetary disk presents a compelling case study, offering a unique perspective on the early stages of planet formation and disk evolution.

The young LkCa~15 (K5, 0.97$M_{\odot}$, [Fe/H]=0.26~dex) \citep{2000ApJ...545.1034S,2021AJ....161..114S} system is a Tauri star located in the Taurus-Auriga star-forming region which is 1-3 Myr old \citep{2019ApJ...877L...3C} and is about 157.19$\pm$0.65 pc away \citep{2023A&A...674A...1G}. LkCa~15 is also claimed to host multiple Jupiter-sized planets \citep{2012ApJ...745....5K,2012ApJ...747..136I,2015Natur.527..342S}, although these claimed planets are much debated. Besides planets, LkCa 15 hosts a protoplanetary disk of radius $\sim$ 160 AU \citep{2019ApJ...877L...3C}, featuring a prominent gap at around 45 to 50 AU. This disk also exhibits multiple substructures, observable both in scattered light and in high-resolution ALMA imaging \citep[e.g.][]{2006A&A...460L..43P,2007ApJ...670L.135E,2008ApJ...682L.125E,2014ApJ...788..129I,2016PASJ...68L...3O,2015ApJ...808L..41T,2016ApJ...828L..17T,2022A&A...663A..23L,2023arXiv231008589R}.

Until the advent of ALMA, optical and near-infrared (NIR) scattered light observations were the best methods to image the protoplanetary disk in high resolution in order to resolve the disk features. Sparse aperture masking interferometry (SAM; \citealt{2006SPIE.6272E.103T}) and AO-assisted pupil stabilised imaging using angular differential imaging (\citealt{2006ApJ...641..556M}) are two major complementary techniques used for obtaining diffraction-limited images from ground-based telescopes. The first proto-planet candidate around LkCa~15 was detected by \cite{2012ApJ...745....5K}. Subsequent investigations by \cite{2015Natur.527..342S} using SAM also reported the presence of three possible protoplanets on Keplerian orbits within 25~au, one of which was recovered in $H_{\alpha}$ (LkCa~15b). However, studies such as \cite{2019ApJ...877L...3C} used high-contrast imaging and suggested that proto-planetary signals detection with SAM are likely inner disk signals. Further, using SAM, \cite{2022ApJ...931....3B} report the detections of two previously observed asymmetric rings at $\sim$17 and $\sim$45 au but found no clear evidence for the  candidate planets. Recently, \cite{2023ApJ...953...55S} have also found that the three companion planet model falls short to explain the positional evolution of the infrared sources as the longer time baseline images lack the coherent orbital motion that would be expected for companions.

\begin{table}[b!]
\caption{\label{t1} Basic stellar parameters for the LkCa~15 system.}
\centering
\begin{tabular}{lcc}
\hline\hline
Parameters & Value & Units \\
\hline
\\
RA (J2000) & 04 39 17.79 & (h m s)  \\
Dec (J2000)  & + 22 21 03.39 & (deg)\\
Distance $^{a}$ &157.19$\pm$0.65 & (pc)\\
$K$-band magnitude & 8.16$\pm$0.018 & (mag) \\

T$_{\rm{eff}}$ $^{b}$   & 4210$^{+185}_{-199}$ & (K)\\
R$_{\star}$  & 1.65 &R$_{\sun}$\\
M$_{\star}$ & 0.97 & M$_{\sun}$ \\
Age   & $\sim$1 & (Myr)\\
\hline
\end{tabular}
\tablefoot{ $^{a}$ \cite{2023A&A...674A...1G} $^{b}$\cite{2021AJ....161..114S}\\
}
\end{table}
Optical and near-infrared (NIR) polarimetric studies have also been conducted on LkCa~15. Using the Zurich Imaging Polarimeter (ZIMPOL), a subsystem of the VLT/SPHERE, \cite{2015ApJ...808L..41T} detected the previously unobserved far side of the disk gap. Later, J$-$band polarimetric observations by \cite{2016ApJ...828L..17T}, using the InfraRed Dual-band Imager and Spectrograph (IRDIS), another VLT/SPHERE subsystem, reported persistent asymmetric structures at the locations of the planetary candidates. \cite{2016PASJ...68L...3O} used H-band polarized intensity images from Subaru/HiCIAO and reported the existence of a bright inner disk misaligned by 13$\pm$4\textdegree with respect to the outer disk. The grain size and polarized intensity fractions were not estimated in these studies.

\begin{table*}
\caption{Observation modes used for LkCa~15 imaging.}
\label{tab:obs_setup}
\centering
\begin{tabular}{l l l c c c c c c c c}
\hline\hline
Arm & UT Date & Mode & DIT (s) & \multicolumn{2}{c}{NDIT} & \multicolumn{2}{c}{On source time} & Seeing (”) & Coherence & Wind  \\
& & & & Sci. & Ref. & Sci. & Ref. & & time (ms) &speed (m/s)\\
\hline
IRDIS & Nov 27, 2020 & \texttt{DP\_0\_BB\_Ks} & 16 & 122 & 20 & 1952 & 320 & 0.69 - 1.37 & 2.8 - 4.5  & 7.05 - 8.18 \\
IRDIS & Dec 8, 2020 & \texttt{DP\_0\_BB\_Ks} & 16 & 312 & 56 & 4992 & 896 & 0.37 - 0.53 & 7.9 - 13.9 & 3.53 - 4.40 \\
\hline
\end{tabular}
\end{table*}

\begin{figure*}[t]
\centering
\includegraphics[width=2\columnwidth]{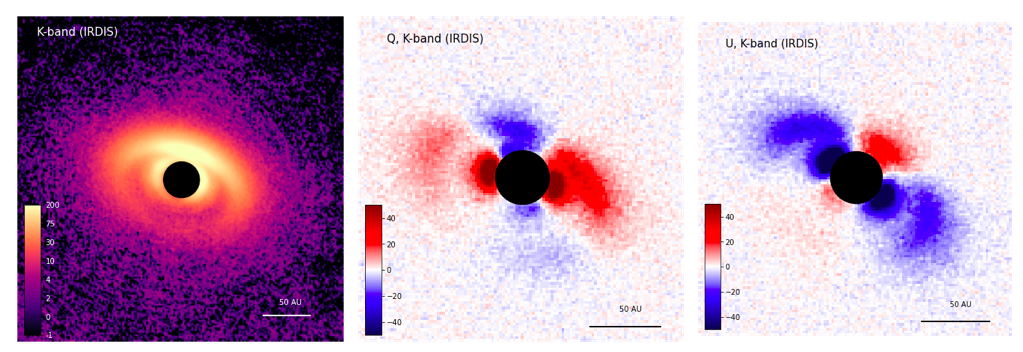}
\caption{Left to right: IRDIS total intensity $K$-band, Stokes $Q-K$-band and Stokes $U-K$-band images as observed from VLT-SPHERE. }
\label{lkca1mg}
\end{figure*} 

With the advent of ALMA, the scattered light images of the LkCa~15 system have been complemented by mm and sub-millimeter images. ALMA images mainly probe the larger grains which cannot be analyzed from the scattered light images. These ALMA images have reconfirmed the existence of a gap at 45 to 50~au which was seen in scattered light images \citep{2006A&A...460L..43P,2007ApJ...670L.135E,2008ApJ...682L.125E,2014ApJ...788..129I,2016PASJ...68L...3O}. Additionally \cite{2019ApJ...881..108J} used the disk density profile to fit the observed gas radial profile of $^{12}$CO obtained using ALMA and found the total disk mass to be 0.1M$_{\odot}$. Although there is no direct evidence for protoplanets detected in ALMA, recent studies in the ALMA 1.3~mm images show multiple gaps and rings. It can be shown using hydrodynamic simulations that the existence of sub-Jovian planets can explain such morphology.Recently, \cite{2022ApJ...937L...1L} presented the deepest dataset on this system and they found that the morphology of the ring at 42 au closely resembles the characteristic horseshoe orbit seen in planet-disk interaction models, with dust accumulation around Lagrangian points L4 and L5 traced by a clump and an arc, respectively.

Even though the presence of companions in LkCa~15 is yet to be established firmly, studies in scattered light and sub-millimeter wavelengths have predicted the likely mass of the proto-planet to be around 6$M_{J}$ \citep{2012ApJ...745....5K}. Other studies such as \cite{2014ApJ...788..129I} using the very large array (VLA) have estimated the mass of the protoplanetary candidate LkCa~15b from the accretion rate and found it to be greater than 5$M_{J}$. Using observations from VLT/SPHERE, Gemini/NICI and Subaru/HiCIAO, \cite{2017ApJ...835..146D} used hydrodynamical simulations in combination with 3D radiative transfer modeling to estimate the mass of the simulated planet in the gap of LkCa~15 to be 0.15 to 1.5 M$_{J}$ depending on the value of viscosity ($\alpha$) in the gap. \cite{2020A&A...639A.121F} using the ALMA 1.3 mm images also performed hydrodynamical simulation and showed that the presence of sub-Jovian planets could explain the observed multi-ringed substructure. Overall, there is a general agreement that the substructures in the LkCa~15 system might have been caused by the presence of a giant planet which is in the process of formation.

Although LkCa~15 has been studied previously in both scattered light and sub-millimeter wavelength, it is important to revisit the LkCa~15 protoplanetary disk because a) New $K_{s}$-band star-hopping data does not have disk self-subtraction,  while previous LkCa~15 observations were post-processed using ADI where self-subtraction \citep{2012A&A...545A.111M,2014ApJ...780...25E,2021A&A...648A..26W} was a major limitation. Star-hopping is a relatively new observing technique with Reference Difference Imaging (RDI) in which one can quickly move the telescope to a nearby star (within 1--2 degrees) and capture its PSF and speckle pattern to use as a reference for subtraction. b) Detect new planets beyond separations of $\sim$ 100~mas ($\sim$ 15.7~au) and c) obtain a self-consistent model that satisfactorily explains the total intensity, polarimetric, and sub-millimeter disk observations. In this paper, we present the self-subtraction free $K_{s}$-band imaging observations of the LkCa~15 protoplanetary disk using the star-hopping technique \citep{2021A&A...648A..26W}. For the first time, the inner 30~au of the disk is clearly visible. We use this observation together with ALMA observations from \cite{2019ApJ...881..108J} and \cite{2020A&A...639A.121F} to create a consistent radiative transfer model of the disk using \texttt{RADMC-3D}. In section~\ref{s2}, we provide a brief description of the star-hopping technique and our observations of LkCa~15. In section~\ref{s3}, we extract the $K_s$-band scattering phase function and polarization fraction from the SPHERE data, and also perform radiative transfer modeling of the protoplanetary disk with \texttt{RADMC-3D}. In section~\ref{s4}, we discuss the possible implications of our disk models and finally, we summarise our results in section~\ref{s5}.

\section{Observation \& data reduction}
\label{s2}
\subsection{LkCa~15 observations}
The new observations of LkCa~15, which is the main focus of this paper, were obtained using the VLT SPHERE instrument \citep{FUSCO2006,2016A&A...587A..58B,2016A&A...587A..57Z,2019A&A...631A.155B}. The SPHERE is a state-of-the-art instrument that incorporates an advanced adaptive optics (AO) system \citep{FUSCO2006,2014SPIE.9148E..1UF} and has three distinct science sub-instruments (1) IRDIS (InfraRed Dual-band Imaging and Spectroscopy) which captures wide field images and performs differential imaging \citep{2008SPIE.7014E..3LD}, (2) IFS (Integral Field Spectrograph) designed for low-resolution spectroscopy, enabling the characterization of exoplanetary atmospheres \citep{2008SPIE.7014E..3EC}, and (3) ZIMPOL (Zurich IMaging POLarimeter), a polarimetric device to detect and study the polarized light scattered by planetary atmospheres and circumstellar disks \citep{2018A&A...619A...9S}. SPHERE can deliver $K$-band Strehl ratio ($\geq$65$\%$) for faint stars like LkCa 15  (G$\sim$12) in median seeing conditions (0.8$\arcsec$-- 1.2$\arcsec$) which makes it one of the best instruments to directly image young planets around faint stars \citep{2022A&A...667A.114J,Jones2025}.

\begin{figure}[t]
\centering
\includegraphics[width=0.8\columnwidth]{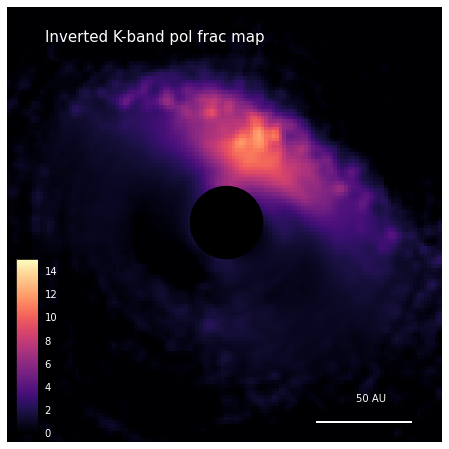}
\caption{Inverted $K$-band polarization fraction (${I_{\text{tot}}}/{\sqrt{Q^2 + U^2}})$ map of LkCa~15 system.  Planets with low polarization, if exist, overlay on a backdrop of highly polarized light coming from the disk, resulting in a robust discrimination caused by a high signal-to-noise ratio (SNR). However, there is no unambiguous planetary signature in this image.A map value of 14 corresponds to a polarization of roughly $1/14\approx7\%$.} 
\label{pmap}
\end{figure}
LkCa~15 was imaged in $K_{s}$-band in IRDIS dual-beam polarimetric imaging (DPI) mode (IRDIS-DPI: \cite{2020A&A...633A..63D,2021A&A...647A..21V}) on two nights, 27 November and 8 December 2020, as part of a larger survey of 29 protoplanetary disks in nearby star-forming regions \citep{2023arXiv231008589R}, aiming to study the disk morphology and detect planets using the star-hopping technique \citep{2021A&A...648A..26W}. The data collected on 27 November was affected by poor observing conditions, as shown in Table~\ref{tab:obs_setup} and the observing sequence was not completed. The data collected on 8 December had a sufficient number of both science frames (312) and reference frames (56) and benefited from much better observing conditions. We only use the latter observations, as adding the lower Strehl-ratio images (effectively lower resolution) from the first observation would degrade the final image quality. For the reference star PSF subtraction, we chose TYC~1279-203-1, which is separated by 10.35' (0.172$^\circ$) from LkCa~15. Moreover, the magnitudes (in $G$) for TYC~1279-203-1 are 11.2 vs. 11.5 for LkCa~15, making it an ideal reference star for PSF subtraction. The science frames were observed in a sequence of $\sim$6 minutes, followed by a hopping overhead time of $\sim$1 minute, after which the reference star was observed for $\sim$2 minutes. The observations were carried out with the \texttt{N\_ALC\_Ks} coronagraph, which has an focal plane mask of radius 120 mas, $\sim$20 au for LkCa 15. In our data analysis, we focus on disk emission detected outside ~150 mas, and therefore, we did not apply a correction for the coronagraphic transmission function in the modeling. The LkCa~15 observations in $K_{s}$-band IRDIS-DPI mode are summarized in Table~\ref{tab:obs_setup}.

In addition to the SPHERE observations, we model ALMA data of LkCa 15 to characterize the structure of the disk. The ALMA Band 7 (880 $\mu$m) data, obtained by \cite{2019ApJ...881..108J}, revealed a dust-depleted cavity with a radius of $\sim$45 au. The surface density profile of the disk follows a power law of the form $\rho \propto r^{-4}$. The ALMA Band 6 (1300 $\mu$m) observations, obtained by \cite{2020A&A...639A.121F}, achieved a resolution of 40 mas $\times$ 60 mas ($\sim$7.5 au), allowing the detection of multiple rings in the disk. Specifically, three rings were identified at $\sim$47, 69, and 100 au, with respective widths of $\sim$9, 6, and 14 au, along with a possible faint outer ring at $\sim$175 au. By modeling ALMA datasets alongside SPHERE images, we capture the spatial distribution of both micron- and millimeter-sized grains. SPHERE observations image scattered light from micron-sized grains, while ALMA data reveal the distribution of millimeter-sized grains and gas beneath them. Together, these datasets enable detailed modeling of the disk's structure and grain distribution.

\subsection{The star-hopping pipeline}
\label{spp}
The basic reduction of the LkCa~15 data was done by the star-hopping pipeline described by \cite{2021A&A...648A..26W}\footnote{\url{https://github.com/zwahhaj/starhopping}}. This pipeline begins with essential preprocessing steps, including flat-field correction, bad pixel removal, and image centering, to produce science and reference frames. We then addressed challenges specific to RDI when there is extended astrophysical signal close to the star. This involved matching the radial intensity profiles of the science and reference PSFs, without which arc-like artifacts appear in the final reduction. Also, we removed any remaining background slopes after sky subtraction. Signal regions were identified using a simple ADI reduction of the referencing images alone, as a detection threshold. This aided in masking signal regions prior to PSF subtraction. For a detailed description of the pipeline's steps, we refer the reader to Appendix~\ref{A4}. We also computed the random and systematic errors introduced by the post-processing steps and found that the typical uncertainties are $\leq$ 4$\%$ (see Appendix~\ref{A3} for more details). 

For the polarimetric reduction, the raw IRDIS exposures were processed through the \texttt{IRDAP} pipeline \citep{2020ascl.soft04015V,2020A&A...633A..63D}, which automatically generates the Stokes \(Q\) and \(U\) cubes by performing polarized flat-fielding, stellar-leakage subtraction, cross-talk correction, and instrumental polarization removal. The final \(Q\), \(U\), \(Q_{\phi}\), and \(U_{\phi}\) images used throughout this work are the direct \texttt{IRDAP} outputs.

\begin{figure}[t]
\centering
\includegraphics[width=1\columnwidth]{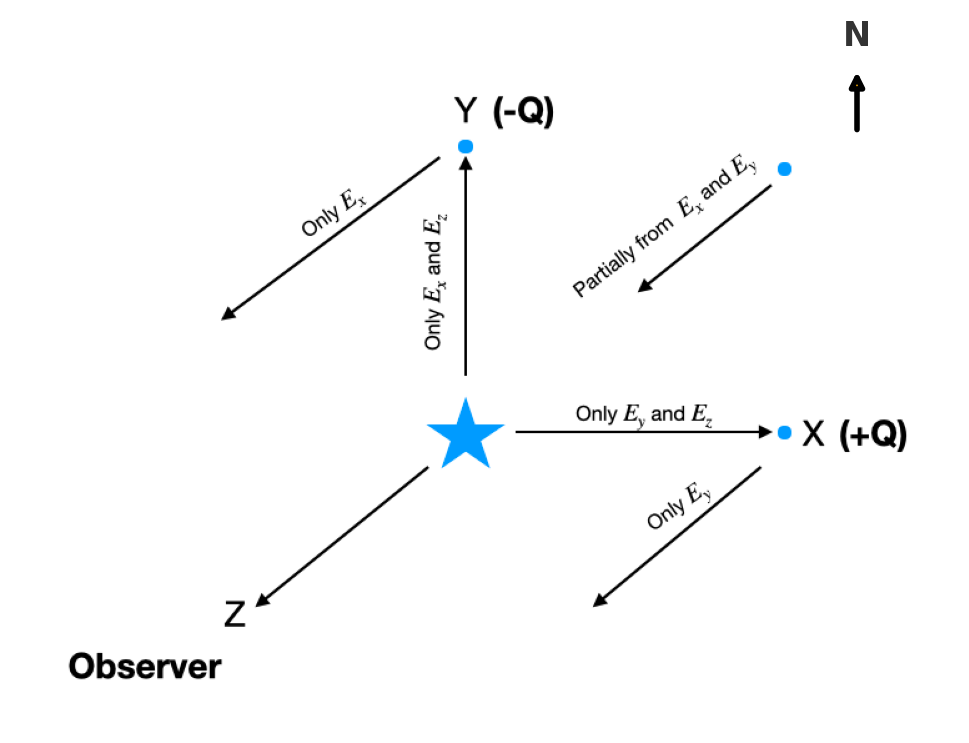}
\caption{Schematic diagram showing the axis orientation for the E-field chosen in this paper. The +Q is aligned along the X-direction and -Q is aligned along the Y-direction. The blue points represent the dust grains, while the arrows indicate the scattered photons towards observer, an incomparably large distance away.}
\label{pol1}
\end{figure} 

\section{Analysis}
In this section, we combine SPHERE/IRDIS \(K_s\)-band scattered light images with ALMA continuum images to obtain a self-consistent model of LkCa~15 that can explain the disk morphology at both NIR and mm wavelengths. We first present the IRDIS data reduction products and construct an inverse polarization-fraction map to search for low-polarization companions. We then fit the disk with \texttt{RADMC-3D} using compact-olivine (Mie) grains in two populations, jointly constrained by IRDIS and ALMA, which sets the global geometry and reproduces the millimeter ring-gap structure. However, this compact-olivine solution leaves systematic mismatches in the \(K_s\) morphology—too little near-side brightness at small-intermediate scattering angles. To diagnose these issues quantitatively, we extract and compare the scattering and polarization phase functions, \(S(\theta)\) and \(p(\theta)\), for the data as well as model them with grains of different porosity and sizes. This analysis confirms the discrepancies and motivates replacing compact spheres with porous CAHP aggregates in the near-IR scattering layer while preserving the jointly constrained geometry. Because millimeter-wavelength optical properties for CAHP are not yet available, this refinement is applied only to the near-IR. We detail the IRDIS flux calibration, and conclude with companion-detection limits derived from disk-subtracted reductions.

\label{s3}
\subsection{IRDIS: Simultaneous polarimetry and total-intensity in $K$-band}
The IRDIS system, a component of SPHERE, has the ability to capture total-intensity and polarised-intensity images in the \(Y\), \(J\), \(H\), and \(K_{s}\) bands. While the \(H\)-band offers superior resolution compared to the \(K_{s}\)-band, the latter demonstrates a markedly $\sim$10\% higher Strehl ratio, especially for faint objects like LkCa~15 (R = 11.61). Consequently, we chose to image the LkCa~15 protoplanetary disk in the \(K_{s}\)-band. Figure~\ref{lkca1mg} shows the total intensity image together with the polarimetric Q and U images while the Figure~\ref{UQphi} (in Appendix \ref{A0}) shows the Q$_{\phi}$ and U$_{\phi}$ images.

From the total intensity and polarization images, we proceeded to estimate the inverse polarimetric fraction map,    \({I_{\text{tot}}/({\sqrt{Q^2 + U^2}})}\), which could potentially reveal planets with low polarization signal. Planetary atmospheres usually scatter light with much lower polarization fraction than disks (also see \cite{2021A&A...647A..21V}). Therefore, in a polarization fraction map, planets would normally appear as intensity dips. Here, we use inverse maps so that planets show up as bright spots on the dark background. To ensure that the map is not dominated by noise in regions with low \(\sqrt{Q^2 + U^2}\), we set these region's pixels to a constant value. This constant was three times the standard deviation (\(\sigma\)) of \(\sqrt{Q^2 + U^2}\) in a background annulus with no significant signal. This was done for regions with \(\sqrt{Q^2 + U^2} < 3\sigma\). However, even with this method, we do not find any strong planetary signatures in the LkCa~15 image shown in Figure~\ref{pmap}. Note that our method can only detect planets if a low-polarization planet is superimposed on a high-polarization disk and detected with a robust SNR.

\subsection{Radiative transfer modeling of the LkCa~15 disk}
We use the radiative transfer code \texttt{RADMC-3D} \citep{2012ascl.soft02015D} to study the properties and structure of the LkCa~15 protoplanetary disk. With \texttt{RADMC-3D}, we are able to generate models of the observed morphology of the disk at different wavelengths of the spectrum, i.e. scattered light with polarimetry at 2.2~$\mu$m (SPHERE data) and thermal emission at sub-milimeter wavelengths (ALMA data). These datasets provide complementary information about the disk, with SPHERE probing the surface layers dominated by micron-sized grains, while ALMA reveals the disk midplane driven by millimeter-sized grains. 

The main goal of this modeling is to create a self-consistent disk model that can explain both scattered light and thermal emission. By constraining key parameters like the disk geometry, grain size distribution, and flaring, we aim to infer the physical structure and composition of the LkCa~15 disk. Our modeling framework involved defining the physical and geometrical properties of the disk and generating synthetic model images that could be directly compared with observations. Through an iterative process of refining the models and comparing them with observational data, we try to find the best-fit solution that align with the observed morphology.

\subsubsection{Model assumptions and parameters}
The radiative transfer model incorporates several physical assumptions about the disk's structure, dust properties, and scattering behavior. Key aspects are outlined below:

\begin{itemize}
    \item \textbf{Scattering theory and dust grain properties:} Dust grains are assumed to follow a power-law size distribution, \(n(a) \propto a^{\zeta}\), with sizes ranging from submicron to millimeter scales. This size distribution ensures that the model captures the effect of both small grains, which dominate scattered light and polarization at shorter wavelengths, and larger grains, which are the primary contributors to sub-millimeter thermal emission. The optical constants for the compact Mie grains are adopted from \cite{1995A&A...300..503D}, while those for the irregular porous grains are taken from \cite{2022A&A...663A..57T}.  
Because we supply pre-computed opacity files for each representative grain size computed using \texttt{OP-TOOL} \citep{2021ascl.soft04010D}, RADMC-3D treats each file as an independent dust species and ignores the \texttt{gsmin} (minimum grain size), \texttt{gsmax}(maximum grain size), and other grain distribution parameters, which only apply when opacities are generated internally from optical constant files (see \cite{2012ascl.soft02015D} for more details). 


    \item \textbf{Disk geometry:} The disk is modeled as an axisymmetric structure defined by its inner and outer radii (\(r_{\mathrm{in}}\), \(r_{\mathrm{out}}\)), inclination (\(i\)), and position angle (\(PA\)). 
\item \textbf{Surface density distribution and gaps:} Each annular zone is assigned an independent power–law surface density, \(\Sigma(r)\propto r^{-plsig}\), between its inner and outer radii (\(R_{\rm in}\), \(R_{\rm out}\)). Where needed, we impose an \emph{explicit} depletion gap using the \texttt{RADMC-3D} controls \texttt{gap\_rin}, \texttt{gap\_rout}, and \texttt{gap\_drfact}. Inside the interval \(r\in[\texttt{gap\_rin},\,\texttt{gap\_rout}]\), the surface density is multiplied by \texttt{gap\_drfact} (\(<1\) for a depletion). Gaps are specified independently for each zone; leaving them unset (or \texttt{gap\_drfact}=1) disables the depletion for that zone.
\item \textbf{Vertical density profile and flaring:}  
  We assume a Gaussian vertical density distribution,  
  \[
    \rho(r,z) \;=\; \rho_0(r)\,\exp\!\Bigl[-\tfrac12\bigl(z/H(r)\bigr)^2\Bigr]\,,
  \]
  where the pressure scale height is  
  \[
    H(r)
      \;=\; H_0\,R_{\mathrm{piv}}\,
            \Bigl(\tfrac{r}{R_{\mathrm{piv}}}\Bigr)^{\mathrm{plh}}
      \;=\; H_0\,r\,
            \Bigl(\tfrac{r}{R_{\mathrm{piv}}}\Bigr)^{\mathrm{plh}-1}\,.
  \]
  Here \(H_0\) is the aspect ratio \(H/r\) at the pivot radius \(R_{\mathrm{piv}}\)\, and `plh' is the flaring exponent.  Thus at \(r = R_{\mathrm{piv}}\) one recovers \(H(R_{\mathrm{piv}}) = H_0\,R_{\mathrm{piv}}\). Although we do not compute a full gas density profile, we use the pressure scale height \(H(r)\) to prescribe the disk's vertical thickness and flaring, thereby defining its three-dimensional geometry for radiative transfer.  For our \texttt{RADMC-3D} models we choose the \(R_{\mathrm{piv}}\) at 50 au.

\end{itemize}


\begin{figure}[t]
\centering
\includegraphics[width=1\columnwidth]{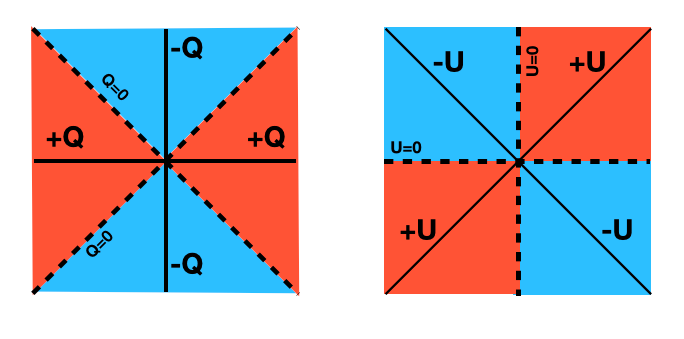}
\caption{The signs of Stokes parameters taken in this paper indicating the positive and negative regions in Q and U images. }
\label{pol2}
\end{figure}

\begin{figure}[t]
\centering
\includegraphics[width=1\columnwidth]{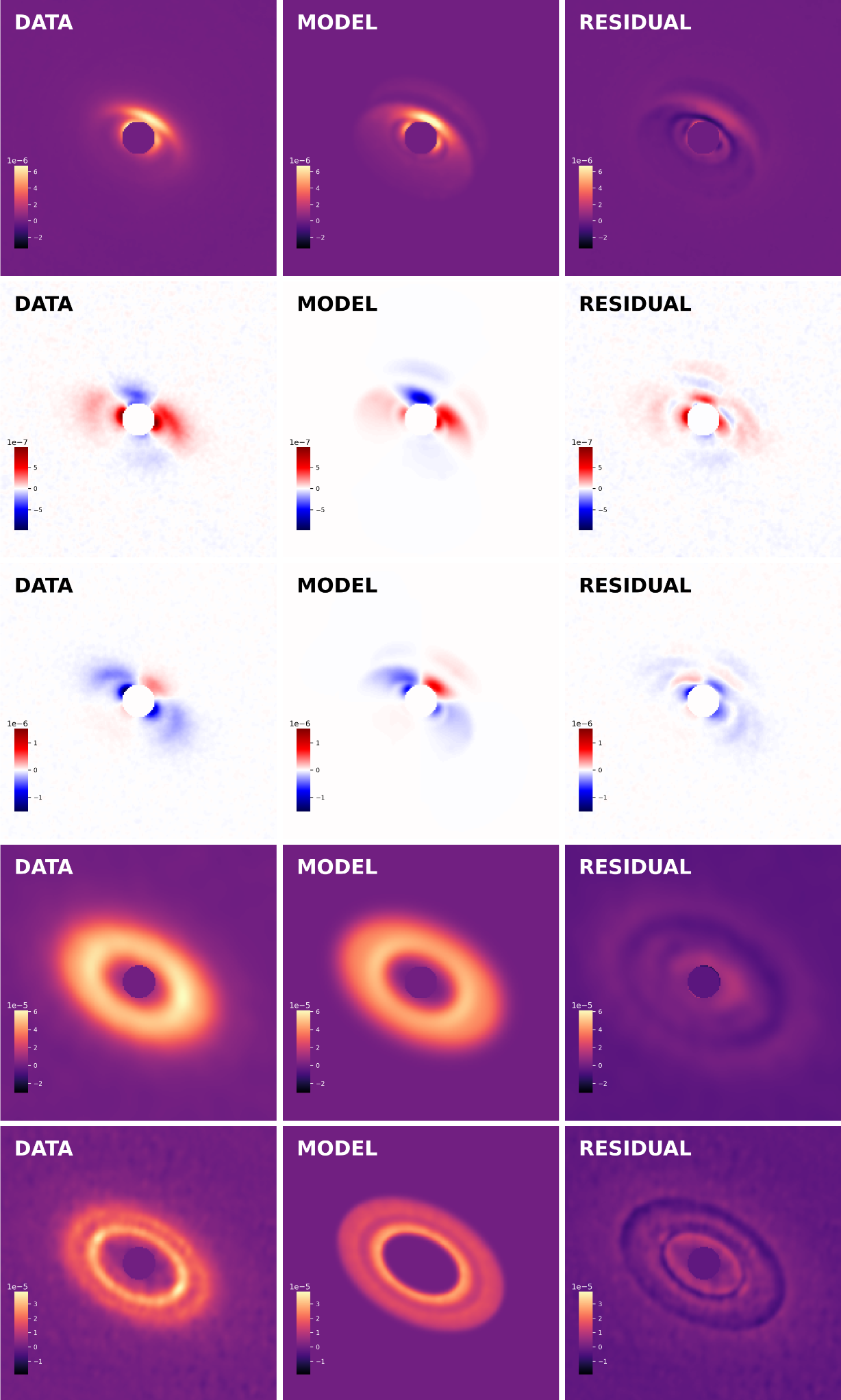}
\caption{Comparison of  \texttt{RADMC-3D} model with Olivine grains and observed data. The sequence from left to right shows the data, the simulated model, and the residuals. The top row is the $K$-band total intensity image, followed by the $K$-band Stokes Q, and the Stokes U. The fourth and fifth rows at the bottom correspond to ALMA $880~\mu$m and $1300~\mu$m images, respectively.}
\label{2dm}
\end{figure}

\begin{figure}[t]
\centering
\includegraphics[width=1\columnwidth]{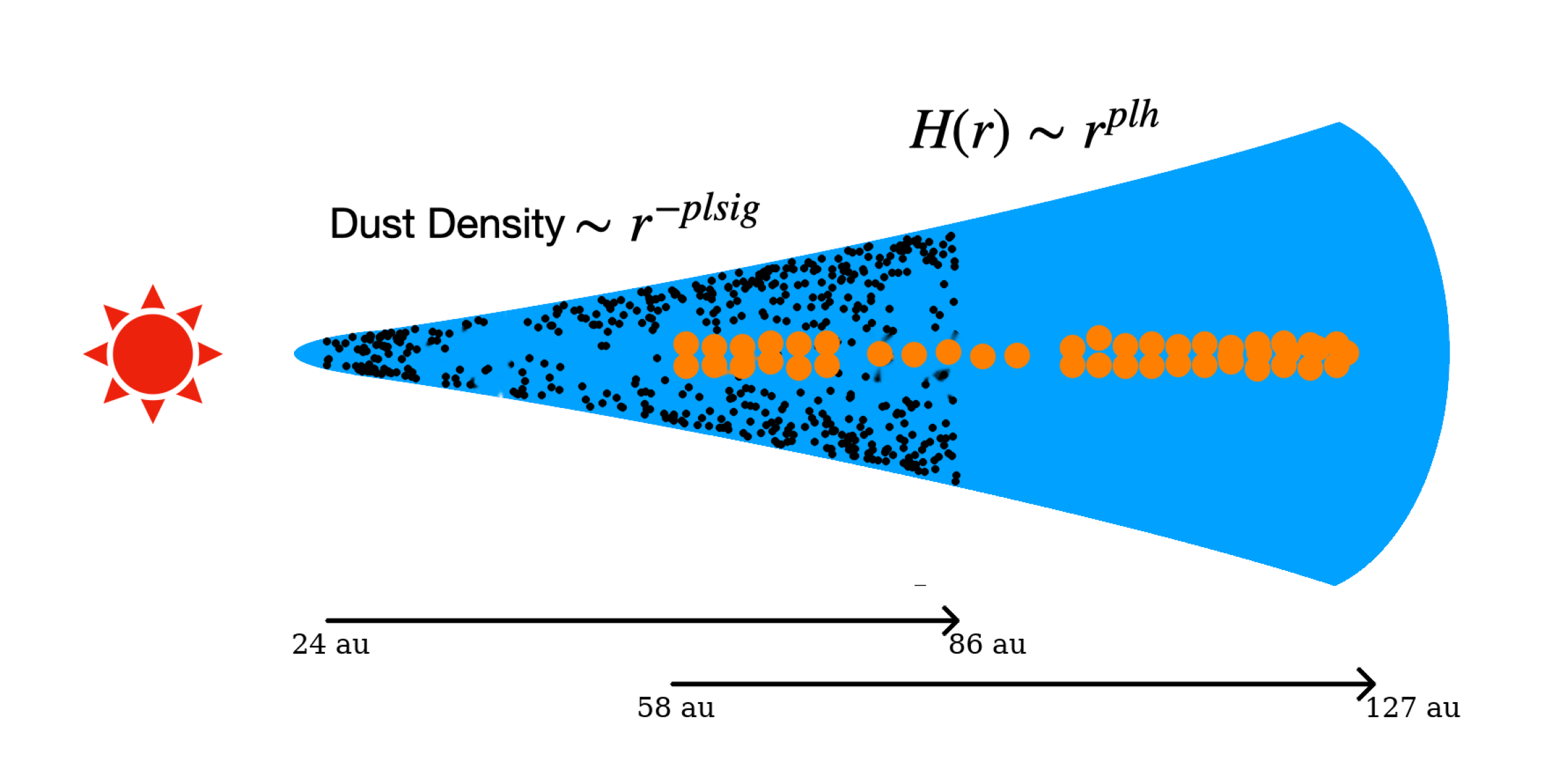}
\caption{Schematic diagram of our LkCa~15 protoplanetary disk with Olivine grains. The black and orange represents the micron and millimeter sized grains. }
\label{sch}
\end{figure}

\subsubsection{The polarimetric images}
\texttt{RADMC-3D} has the capability to generate polarimetric images (Stokes: Q, U) along with the intensity images. Because the axes used to define the components of the electric field vectors are arbitrary, the appearance of the Q and U images also varies depending on the chosen convention. To make the images obtained from \texttt{IRDAP} and \texttt{RADMC-3D} consistent with each other, for both images we define the axis of E-field relative to the North as shown in Figure~\ref{pol1}. The positive Q (+Q) is oriented along the X-direction, whereas the negative Q (-Q) is aligned along the Y-direction. A photon emitted from the star, as an electromagnetic wave, has no electric field component along its direction of propagation. Consequently, the electric field is confined to the two orthogonal directions. If a photon approaches the particle along the Y-direction, its electric field will have $E_{x}$ and $E_{z}$ components. After scattering towards the observer along the Z-direction, only the $E_{x}$ component will be preserved. Similarly, if the photon approaches  along the  X-direction, then only $E_{y}$ will be preserved. At  intermediate angles, both $E_{x}$ and $E_{y}$ components contribute, with Q$>$0 if $E_{x}$>E$_{y}$, and Q$<$0 if $E_{x}$<$E_{y}$. For an inclined disk, the orientations of the $E_{x}$ and $E_{y}$ components vary, resulting in uneven positive regions (red) in the east-west direction and negative regions (blue) in the north-south direction. 

Similar to the Q images, the U images also exhibit positive and negative regions, which are represented in a similar manner as illustrated in Figure~\ref{pol2}.  Since the orientation of the +Q and -Q is arbitrary, different models and reduction tools may use different directions to represent  +Q and -Q. Therefore, it is necessary to apply a suitable corrections before comparing the images. For example, the orientation of the images obtained from \texttt{IRDAP} pipeline and \texttt{RADMC-3D} has an offset of +90$^o$ clockwise, and thus a proper alignment is necessary before comparison. The second and third row in Figure~\ref{2dm} shows both the observed and modeled polarimetric images for the LkCa~15 after applying the offset correction. We find that the orientation of the positive and negative regions of the Q and U image and the model are correct though there is a considerable residual for both cases. A circular mask of radius $\sim$ 20 au has been applied both to the data and the model to block the central PSF which is necessary for the scattered light images. Since we mask the inner $\sim$20\,au of the disk, we exclude regions where PSF convolution could artificially alter the polarization signal from unresolved sub-FWHM structures. The SPHERE/IRDIS \(K_s\)-band PSF has a FWHM of $\sim$50\,mas (corresponding to $\sim$7.85\,au at 157\,pc), so our mask spans nearly three times the PSF width. As a result, small-scale polarimetric features susceptible to PSF-induced artifacts \citep{2019A&A...627A.156H,2021A&A...655A..37T,2024A&A...683A..18M} are already excluded from both the modeling and the $\chi^2$ computation, thereby justifying our approach.

\subsubsection{Flux Calibration}
We converted the IRDIS images in ADU units to flux density in Jansky (Jy) using the known $K$-band flux density of LkCa 15 (\(F_\star = 0.369 \, \mathrm{Jy}\); \cite{2003yCat.2246....0C}). First, we performed aperture photometry on the non-coronagraphic frames to measure the stellar count rate. We then scaled this by the filter transmission accounting for the neutral density (ND) filter attenuation and by the exposure time (DIT) to establish a conversion factor from ADU to Jy, \(\mathrm{CPS}_\star\) (in ADU s$^{-1}$). Finally, the coronagraphic disk flux (\(F_\mathrm{disk} \, [\mathrm{Jy}]\)) was computed as,

\begin{equation}
F_\mathrm{disk} = \frac{\mathrm{ADU}_\mathrm{disk}}{\mathrm{DIT}_\mathrm{disk}} \times \frac{F_\star}{\mathrm{CPS}_\star}.
\end{equation}

Where \(\mathrm{ADU}_\mathrm{disk}\) is the disk signal in detector counts (after subtracting background and stellar PSF), and \(\mathrm{DIT}_\mathrm{disk}\) is the Detector Integration Time (DIT) or the exposure time of a single frame.

\subsubsection{Radiative Transfer Modeling with Compact Olivine Mie Spheres}\label{sec:rt_mie}
To model the protoplanetary disk in scattered light and at sub-millimeter wavelengths, we adopt a minimal two-population dust model with micron-sized grains in the surface layer and millimeter-sized grains in the mid-plane. In the ALMA images, the disk appears geometrically thin because the thermal continuum is dominated by large grains that have settled toward the mid-plane \citep{2016ApJ...816...25P}. In contrast, the SPHERE near-infrared images trace a significantly flared, three-dimensional scattering surface (Fig.~\ref{2dm}), which requires small grains. A single grain population cannot reproduce both the flat mm continuum and the flared NIR morphology, which motivates our two-component dust prescription.

As discussed in Sect.~\ref{s2}, we model the LkCa~15 protoplanetary disk using SPHERE $K_s$-band (2.2~$\mu$m) Stokes $I$, $Q$, and $U$ images together with the ALMA 880~$\mu$m and 1300~$\mu$m continuum maps. The model images are first scaled to the observed images by minimising the rms difference in an annulus between 10 and 75 pixels. We then evaluate a mean reduced image chi-square, $\overline{\chi}_{R,\mathrm{img}}^2$, which measures the agreement in morphology after this renormalisation, and a separate flux chi-square, $\overline{\chi}_{f}^2$, which compares the integrated fluxes in the NIR and ALMA bands (see Appendix~\ref{A1}). The total reduced chi-square is defined as $\overline{\chi}_{R}^2 = \overline{\chi}_{R,\mathrm{img}}^2 + \overline{\chi}_{f}^2$, but in practice we first search for good solutions by minimising $\overline{\chi}_{R,\mathrm{img}}^2$ alone, as this is computationally faster; we then use $\overline{\chi}_{R}^2$ to refine the fit once an acceptable morphological match has been found.

For the NIR total disk flux we adopt a conservative 15\% uncertainty, following \citet{2024A&A...687A.257W}. This budget accounts for the dominant SPHERE systematics, namely absolute photometric calibration uncertainties ($\sim5\mbox{--}10\%$), neutral-density filter transmission ($\sim3\%$), flat-field and background-subtraction residuals ($\sim3\%$), and star-hopping PSF-subtraction uncertainties ($\sim5\mbox{--}10\%$) \citep{SPHERE_Documentation}. For the ALMA 880~$\mu$m and 1.3~mm data we assume flux uncertainties of order 10\%, so that the mm continuum provides much tighter constraints on the total flux than the NIR images. The resulting values of $\overline{\chi}_{R,\mathrm{img}}^2$ and $\overline{\chi}_{f}^2$ for our best-fit model are listed in Table~\ref{chitab}.

\begin{table}
\caption{\label{bf2} The best-fit parameters using the Olivine grains.}
\centering
\begin{tabular}{lcll}
\hline\hline
Parameters &  Values & LB & UB \\
\hline
$inc$ [deg] & -50.15 & -51.11 & -49.78\\
$posang$ [deg] & 61.42 & 61.36 & 62.57 \\
\hline 
\hspace{65pt}$\mu$m region \\
\hline
$gs$ [in $\mu$m]  & 12  & 10 & 16 \\
$M_{dust}$ [$M_{\odot}$]  & 4x10$^{-5}$ &  3x10$^{-5}$ &  5x10$^{-5}$\\
$R_{in}$ [au] & 24 & 21 & 29  \\
$R_{out}$ [au] & 86 & 81 & 92  \\
$plh$ & 1.42 & 1.32& 1.48 \\
$H_0^{a}$ & 0.08  & 0.08 & 0.09\\
$plsig$ & 0.76 & 0.72 & 0.84 \\
$gap_{\rm in}$ [au] & 36 & 34 & 39\\
$gap_{\rm out}$ [au] & 42 & 39 & 44\\
$gap_{\rm drfact}$ & 0.12 & 0.04 & 0.21\\
\hline
\hspace{65pt} MM region \\
\hline
$gs$ [in $\mu$m]  & 2000  & 1500 & 2200 \\
$M_{dust}$ [$M_{\odot}$]  & 1.5x10$^{-3}$ & 1.3x10$^{-3}$ & 1.8x10$^{-3}$ \\
$R_{in}$ [au] & 58 & 55 & 62 \\
$R_{out}$ [au] & 127 & 124 & 133  \\
$plh$ & 1.33 & 1.24 & 1.39  \\
$H_0^{a}$  & 0.006 & 0.005 & 0.007\\
$plsig$ &  -0.34 & -0.43 & -0.24  \\
$gap_{\rm in}$ [au] & 77 & 76 & 80\\
$gap_{\rm out}$ [au] & 96 & 94 & 100\\
$gap_{\rm drfact}$ & 0.34 & 0.30 & 0.39\\
\hline
\end{tabular}
\tablefoot{The UB and LB represents the Upper and lower bounds for the estimated parameters. For more details on how UB and LB are calculated, see section~\ref{cmp}.\\$^{a}$ $R_{\mathrm{piv}}$ =  50 au
}
\end{table}

\begin{figure}
\label{scatt}
  \centering
    \centering
    \includegraphics[width=\linewidth]{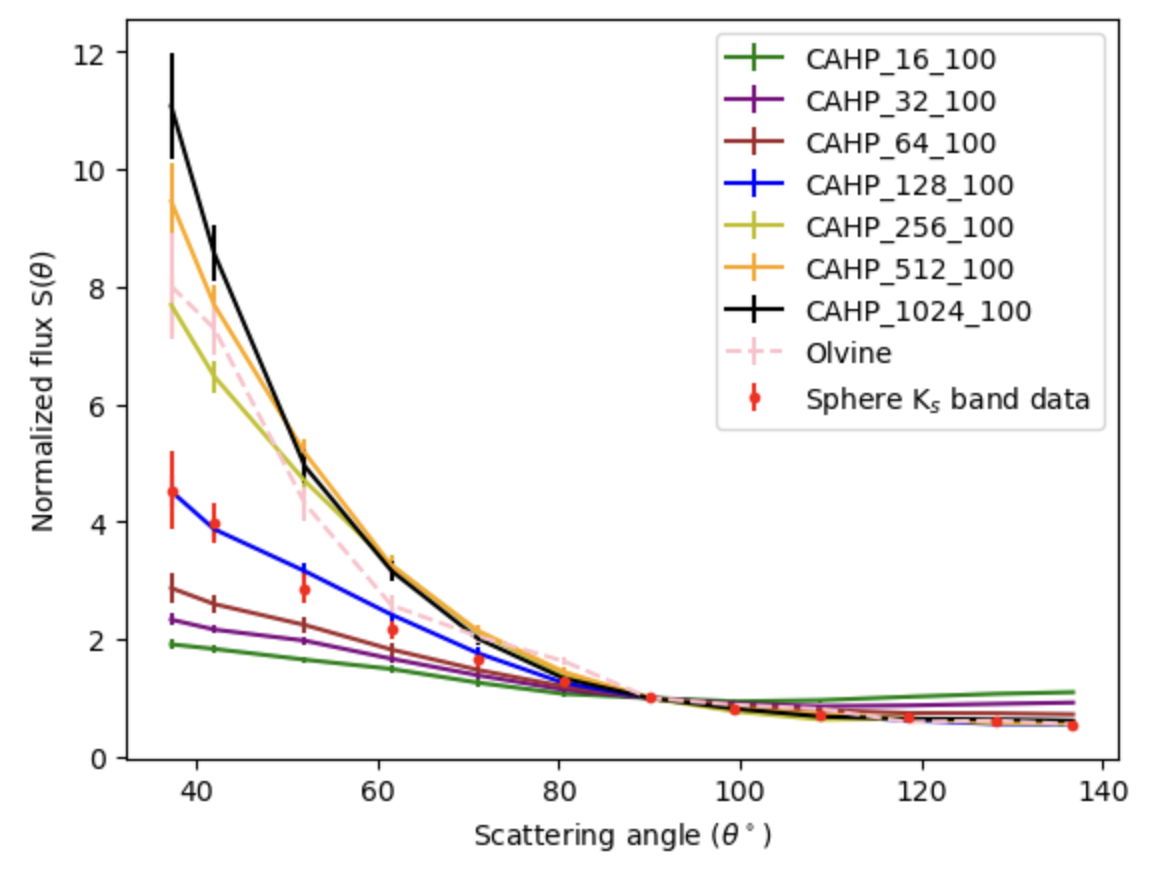}
    

  \vspace{1ex}

    \centering
    \includegraphics[width=\linewidth]{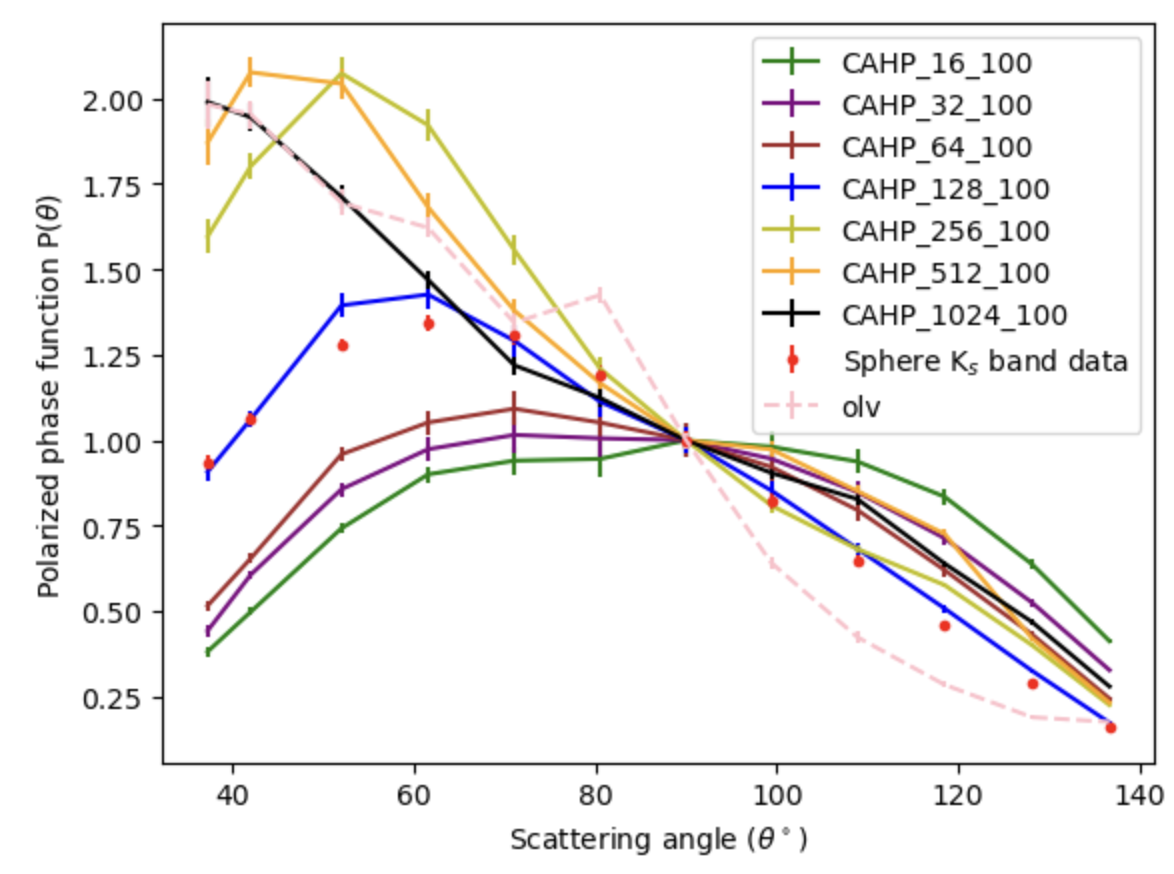}

  \caption{Comparison of the SPHERE Ks-band scattered light phase function (red points) to a family of CAHP aggregate grain models with monomer counts 16, 32, … 1024 (coloured solid lines), and to a compact $\sim$ 8$\mu$m olivine grain (pink dashed). \textbf{Top:} Normalized scattered-light intensity (flux) as a function of scattering angle. \textbf{Bottom:} Polarized phase function versus scattering angle for the same model series and data points.}
  \label{fig:aggregate-scattering}
\end{figure}

Our best-fit olivine model (Fig.~\ref{2dm}) consists of a micron-sized surface layer extending from $\sim$ 25~au to 85~au and a millimetre-emitting ring from $\sim$ 55~au to 130~au, with a prominent gap between $\sim$ 75~au and 100~au in the latter. The overall disk geometry is well reproduced, with an inclination of \(\mathrm{inc} \approx -50^\circ\) and a position angle of \(\mathrm{PA} \approx 61.5^\circ\). Following \citet{1995A&A...300..503D}, we adopt amorphous olivine with a 50:50 forsterite-fayalite composition (equal fractions of Mg$_2$SiO$_4$ and Fe$_2$SiO$_4$). The corresponding best-fit parameters for the micron and millimetre components are listed in Table~\ref{bf2}, and a schematic view of the resulting disk structure is shown in Fig.~\ref{sch}.

The two representative grain sizes used in this model (12~$\mu$m and 2~mm) were not chosen \emph{a priori} but treated as free parameters in an iterative $\chi^{2}$ minimisation with \texttt{RADMC-3D}. Grain sizes $\sim$ 12~$\mu$m provide the best match between total intensity and polarization in the SPHERE $K_s$-band, reproducing both the moderately forward-scattering phase function and the observed surface-brightness distribution; in contrast, sub-micron grains yield polarized intensities that are too low, while larger ($>20~\mu$m) grains produce phase functions that are overly peaked at small scattering angles. For the ALMA 880~$\mu$m and 1.3~mm continuum, the observed fluxes and ring morphology are best matched by grains of order 1--2~mm, with a best fit size of $\sim$2~mm that minimises the residuals. These two grain sizes should therefore be regarded as effective representatives of the dominant scattering and emitting populations at the disk surface and in the mm-emitting mid-plane, respectively.

Despite reproducing the overall geometry and the ring-gap structure seen in the ALMA images, the compact-olivine model shows clear shortcomings in the SPHERE $K_s$-band scattered light. The residuals in Fig.~\ref{2dm} reveal that the inner disk in total intensity model is too strongly forward-scattered compared to the observations, and the spatial distribution of the Stokes $Q$ and $U$ lobes are not well reproduced. In particular, the observed polarized intensity shows more emission at large scattering angles (on the far side of the disk) than our compact 12~$\mu$m Mie grains can generate. Because these grains produce a highly forward-peaked phase function with relatively weak backward scattering, they are unable to match the detailed polarimetric structure of the SPHERE data even though they reproduce the ALMA continuum rings. This, in turn, motivates the exploration of alternative dust models, which we present in the next subsection; there we show that \texttt{CAHP-128-100},nm grains yield the closest agreement with the observed polarization phase curves of the LkCa~15 disk.

\begin{figure}[t]
\centering
\includegraphics[width=1\columnwidth]{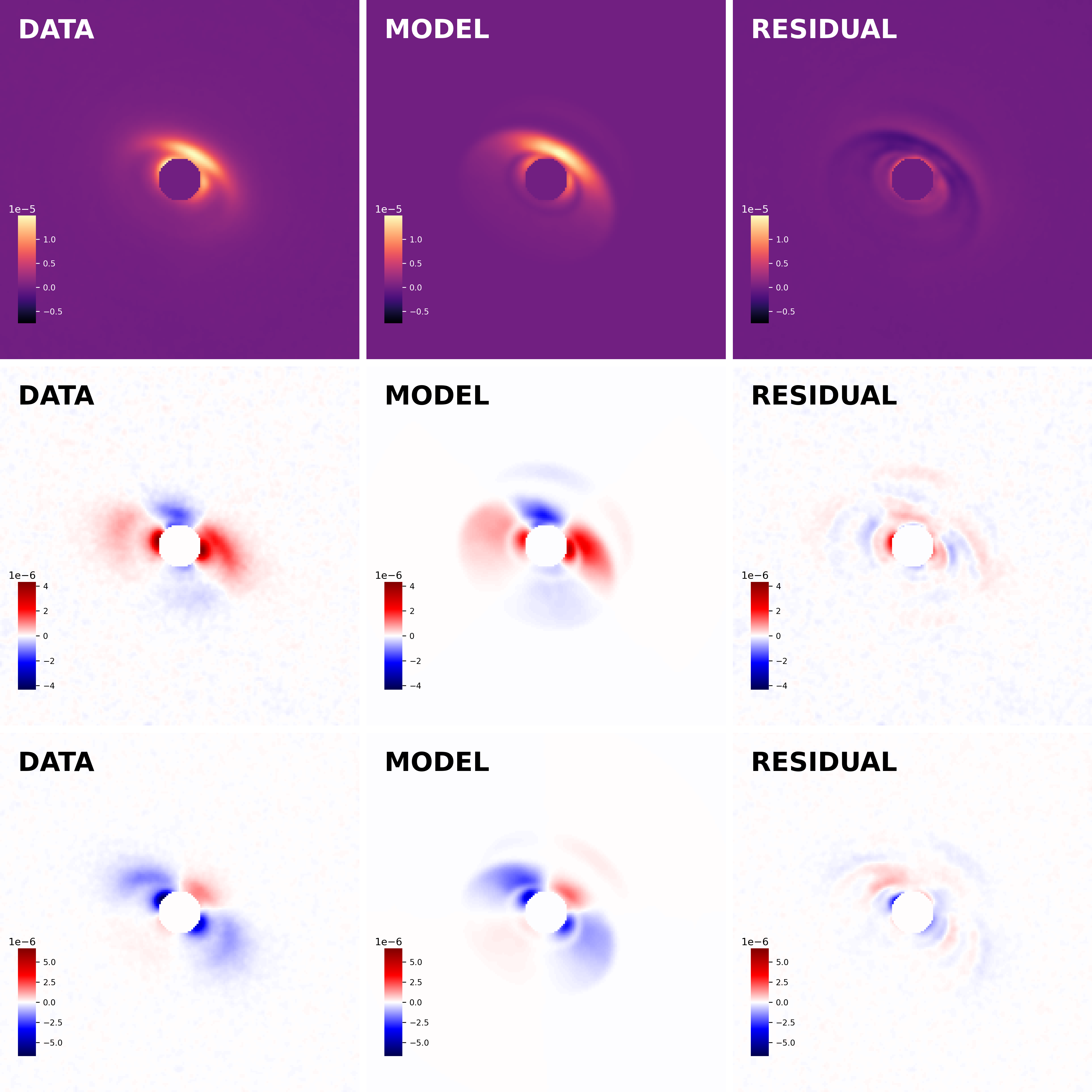}
\caption{Side by side comparison between the observations and a \texttt{RADMC-3D} model with \texttt{CAHP-128-100}nm grains, plus residuals. Left→right: data, model, residuals. Rows (top to bottom): $K$-band total intensity, Stokes $Q$, and Stokes $U$. We didn't model the ALMA images using \texttt{CAHP} grains as we don't have the optical properties of the \texttt{CAHP} grains at mm wavelengths.}
\label{3dm}
\end{figure}
\subsubsection{Scattering Phase Function and Polarization Diagnostics} \label{spp} To constrain the size and structure of the dust grains responsible for near-infrared scattering in the LkCa 15 disk, we used the \texttt{diskmap}\footnote{https://diskmap.readthedocs.io/en/latest/modules.html} \citep{2016A&A...596A..70S} to extract both the scattering phase function and polarization fraction from our $K_s$-band observations at a deprojected radius between 35 to 70 au. These diagnostics directly characterize the angular distribution of scattered light and the degree of polarization as functions of scattering angle, bypassing the complexities of absolute flux calibration and multi-zone radiative transfer. The \texttt{diskmap} tool maps each pixel in our scattered-light images to a corresponding scattering angle, based on the three-dimensional geometry of the LkCa~15 disk. This enabled us to construct the total intensity phase function, $S(\theta)$, and polarization fraction, $P(\theta)$, as a function of the scattering angle $\theta$. For comparison with theoretical models, we normalized the phase function such that $flux=1$, setting the intensity at $\theta=90^\circ$ as a reference point. From the normalized intensity flux plot as shown in Figure~\ref{fig:aggregate-scattering}, our analysis reveals that $S(\theta)$ shows a moderate forward-scattering enhancement, increasing by a factor of $\sim5$ from $90^\circ$ to $35^\circ$. This behavior is more forward-peaked than for ideal Rayleigh scatterers (sub-micron grains), but less extreme than expected for compact, micron-sized spheres, which would produce a much sharper forward-scattering peak as seen in the case of olivine. From the $P(\theta)$ plot in Figure~\ref{fig:aggregate-scattering}, the observed data points (red) show a broad peak near $\theta\sim70^\circ$ and decline at both smaller and larger angles, with an amplitude and shape that cannot be reproduced by compact olivine grains or pure Rayleigh scatterers.

\begin{figure}[t]
\centering
\includegraphics[width=1.0\columnwidth]{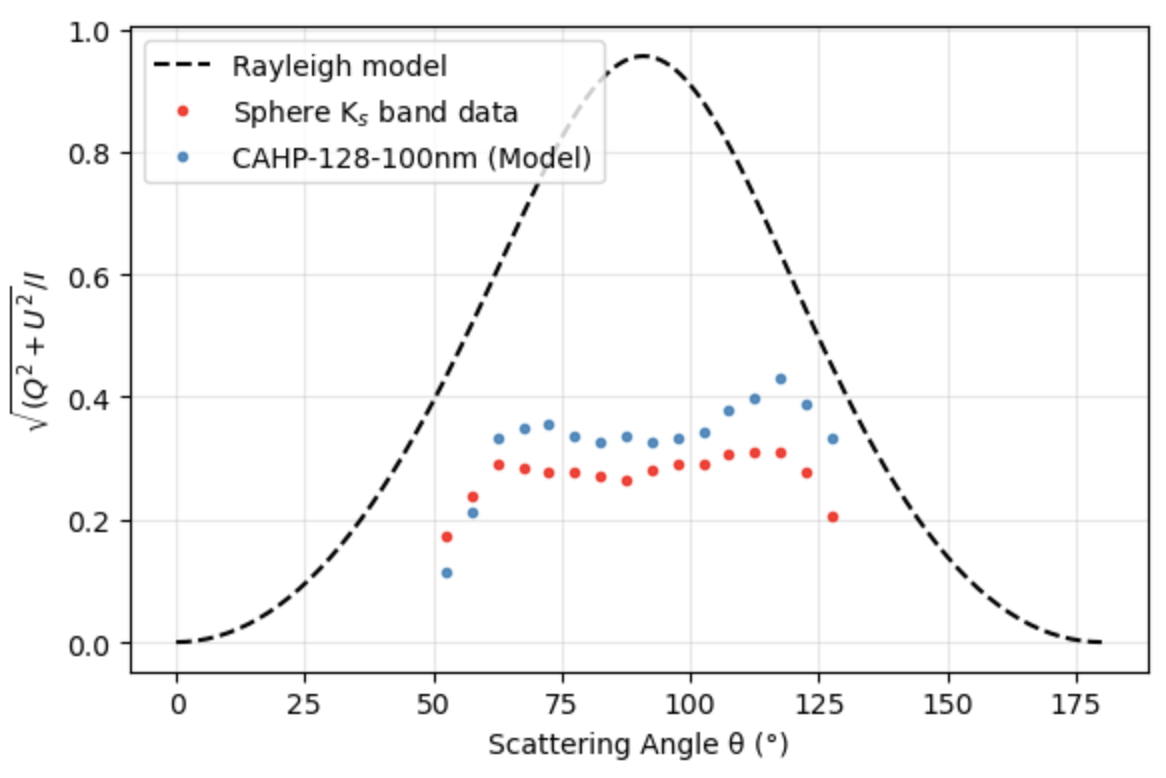}
\caption{Polarization fraction as a function of scattering angle for the LkCa~15 disk. Red points: SPHERE $K_s$-band data (debiased); blue: CAHP-128-100nm high-porosity aggregate model; black dashed: Rayleigh limit. The broad, sub-Rayleigh peak indicates that scattering is dominated by porous aggregate grains rather than compact Mie spheres.}
\label{pol_fracc}
\end{figure}

To interpret these findings, we compared the extracted phase curves to theoretical models generated with \texttt{AggScatVIR}\footnote{https://github.com/rtazaki1205/AggScatVIR/} \citep{2022A&A...663A..57T,2023ApJ...944L..43T}, which simulates scattering by porous aggregate grains composed of sub-micron monomer clusters. We tested several classes of aggregates such as Fractal aggregates \texttt{(FA)} and compact aggregates \texttt{(CA)}, with low porosity (\texttt{LP}), medium porosity (\texttt{MP}), and high porosity (\texttt{HP}) models and varying the number of monomers and grain size. Among these, the CAHP$-$128$-$100nm model, representing an aggregate of 128 monomers (each with 0.1 $\mu$m radius) and a porosity of $\sim$87\%, provided the closest match to the observed phase function and polarization fraction as seen in Figure~\ref{fig:aggregate-scattering}. This model successfully reproduces both the moderate forward-scattering enhancement in $S(\theta)$ and the reduced, flattened polarization peak in $P(\theta)$. The polarization fraction, $P=\sqrt{(U^2+Q^2)}/I$, displays a broad maximum near $\theta\sim90^\circ$, with a peak value $\sim$ 0.35, and decreases at smaller and larger angles. This peak is significantly lower than the Rayleigh limit ($P(\theta)=1$) as shown in Figure~\ref{pol_fracc}, implying that the grains responsible for the observed scattering are outside the Rayleigh regime and/or possess irregular, aggregate structures. By combining disk geometry derived from \texttt{RADMC-3D} with the \texttt{diskmap} based phase extraction and aggregate grain modeling using \texttt{AggScatVIR}, our analysis provides robust, independent confirmation that the scattering surface of LkCa 15 is dominated by porous, micron-scale aggregates with moderate anisotropy and polarization efficiency. Thus we decided to perform our RT modelling for our disk with \texttt{CAHP-128-100}\,nm grains.

\subsubsection{Radiative Transfer Modeling with \texttt{CAHP} Grains}
\label{cahp}

The compact olivine Mie model does not reproduce the observed $K_s$-band morphology as it fails to match the near-side brightness at small scattering angles in Stokes $I$ and yields ansae with too low contrast in Stokes $Q$ and $U$ (Fig.~\ref{2dm}). To test whether this mismatch arises from the scattering physics of the surface grains, we replace compact spheres in the NIR scattering layer with irregular porous aggregates and recompute the models under identical assumptions for geometry, grid, wavelength sampling, image masks, and PSF treatment. Following the analysis in Sect.~\ref{spp}, we adopt \texttt{CAHP-128-100}\,nm grains, which provide the best match to the measured scattering phase function and polarization fraction, and we ingest the corresponding scattering matrices into \texttt{RADMC-3D}. We then perform the same $\chi^2$ minimisation procedure as in Sect.~\ref{sec:rt_mie}. The best-fit CAHP model preserves the global geometry, with an inclination of about $49^\circ$ and a position angle of $61^\circ$. The NIR scattering layer extends from $\sim$ 25 to 85~au with modest flaring ($\mathrm{plh}=1.17$) and $H/R \approx 0.08$ at 50~au. A shallow gap is recovered from $\sim$ 36 to 42~au with a depletion factor of 0.41, and the dust mass in the surface-layer is $2\times10^{-5}\,M_\odot$ (Table~\ref{bf3}).

With \texttt{CAHP} aggregates the $K_s$-band morphology improves in both total intensity and polarimetry. The near-side inner arc becomes brighter and more structured, reducing the residuals seen with compact olivine spheres, and the ansae in Stokes $Q$ and $U$ become sharper, allowing us to reproduce the ``ear''-like lobes seen in the SPHERE data. Quantitatively, the mean reduced $\chi^2$ for the scattered-light images decreases from 1.58 in the compact-olivine model to 1.34 in the \texttt{CAHP} case, and the individual reduced $\chi^2$ values for the NIR images improve from 1.36, 1.45, and 1.93 (Stokes $I$, $Q$, and $U$) to 1.06, 1.24, and 1.74, respectively. The fractional flux differences likewise shrink from 14/34/43\% to 11/8/3\% for $I/Q/U$ (Table~\ref{chitab}). We therefore conclude that irregular porous aggregates provide a significantly better description of the NIR scattering layer than compact Mie spheres. These improvements, however, apply only to the SPHERE $K_s$-band scattered light, as the ALMA 880~$\mu$m and 1.3~mm images are not modelled with \texttt{CAHP} grains.

This restriction reflects the current limitations of the \texttt{CAHP} optical-property libraries, they provide validated scattering matrices for sub-micron to micron-sized aggregates at near-infrared wavelengths, but do not supply absorption and scattering opacities or full Mueller matrices for grains with sizes $\gtrsim 3.78\,\mu$m. As a result, we cannot self-consistently model the ALMA 880 μm and 1.3 mm continuum, which is dominated by absorption from settled millimetre-sized grains.

Finally, our radiative-transfer modelling is affected by parameter degeneracies, particularly between the pressure scale height ($H_0$), the flaring index ($plh$), and the surface-density slope ($plsig$). These quantities are correlated: changes in one can often be compensated by adjustments in the others. For example, increasing $H_0$ can be offset by decreasing $plh$ or increasing $plsig$, producing very similar apparent disk morphologies, and analogous trade-offs exist between $plh$ and $plsig$. As a result, the best-fit values of these structural parameters are not uniquely constrained and should be interpreted with caution when examining the detailed numbers in Tables~\ref{bf2} and~\ref{bf3}.

\begin{table}
\caption{\label{bf3} The best-fit parameters using the CAHP-128-100nm dust grains.}
\centering
\begin{tabular}{lcll}
\hline\hline
Parameters &  Values & LB & UB \\
\hline
$inc$ [deg] & -49.22 & -50.31 & -49.02\\
$posang$ [deg] & 60.89& 60.32  & 61.42 \\
$M_{dust}$ [$M_{\odot}$]  & 2x10$^{-5}$ &  1.5x10$^{-5}$ &  2.2x10$^{-5}$\\
$R_{in}$ [au] & 23 & 21 & 24 \\
$R_{out}$ [au] & 86 & 83 & 88  \\
$plh$ & 1.17 & 1.15& 1.20 \\
$H_0^{a}$ & 0.085  & 0.082 & 0.087\\
$plsig$ & 0.03 & 0.05 & 0.01 \\
$gap_{\rm in}$ [au] & 36 & 34 & 39\\
$gap_{\rm out}$ [au] & 42 & 39 & 44\\
$gap_{\rm drfact}$ & 0.41 & 0.36 & 0.45\\
\hline
\end{tabular}
\tablefoot{$^{a}$ $R_{\mathrm{piv}}$ = 50 au. The ALMA continuum is not fitted with \texttt{CAHP} grains, as their mm-wavelength opacities are unavailable.\\
}
\end{table}

\subsubsection{Constraints on model parameters}
\label{cmp}
Determining the uncertainty of our model parameters is a challenging task due to the high computational cost of Monte Carlo Bayesian estimations. Each model takes approximately one minute to compute, mainly because of the high resolution of our dataset. A representative MCMC run with 50 walkers and 1000 iterations would therefore require of order $\sim$1000 hours. Using shorter runs does not provide reliable constraints on our parameters, as shown in Appendix~D of \cite{2024A&A...687A.257W}. To address this issue, we adopted the methodology proposed by \cite{2024A&A...687A.257W} to obtain an approximate estimate of the uncertainties near the local minimum. For each parameter, we calculated the best chi-squared ($\chi^2$) along a linear grid centered around the optimal fit point, allowing the other parameters to vary. To determine the uncertainty for each parameter, we choose the interval along the grid where the variation in $\chi^2$ remains within 1$\%$ of its minimum value. We chose this threshold because our $\chi^2$ estimates had a precision of around 0.5$\%$. It's important to note that this method doesn't find the global best-fit model or provide definitive uncertainties. However, it does effectively demonstrate the relative strength or weakness of the constraints around our best-fit model.

The lower and upper bounds for our models, as shown in Table~\ref{bf2} and Table~\ref{bf3} reveal that most parameters are tightly constrained at the local minimum. The inner and outer radii ($R_{in}$ and $R_{out}$) critically affect the extent of the disk emission in scattered light and thus are strongly constrained. The dust mass ($M_{dust}$) is well-constrained, as it strongly affects the optical thickness in the Ks band. The power law height ($plh$) and height ratio disk ($H_0$) are moderately constrained. They show evidence for some vertical thickness observed in scattered light and less in thermal emission. The mm zone larger grains have a less pronounced effect on the observed scattered light, they significantly impact the thermal emission seen in the ALMA bands. The dust mass ($M_{dust}$) in this region is somewhat constrained as it directly affects the optical depth. The outer radius ($R_{out}$) as well as the gap location is also well-constrained, as it is crucial for defining the spatial extent observed in ALMA images. Overall, parameters such as dust mass and vertical height, which directly influence the 3D appearance, the clarity of the gap, and the overall brightness of the disk, are better constrained in the submicron and micron zones. In contrast, $plsig$, which influences the mm zone brightness, mainly in ALMA imaging, is less tightly constrained.

\subsection{Companion detection limit}
\label{sec_det_lim}
\begin{figure}[t]
\centering
\includegraphics[width=1.0\columnwidth]{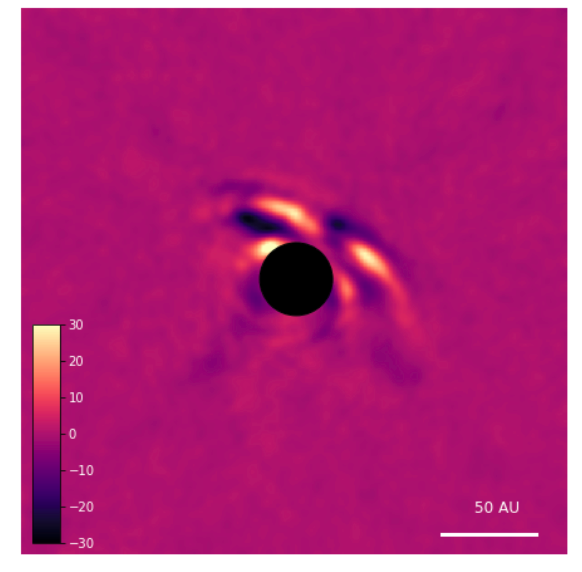}
\caption{\textbf{Left:} IRDIS $K$-band reduction with \texttt{LOCI} algorithm after removing most of the disk in order to detect obscured planets and to estimate the contrast limit as shown in Figure~\ref{cc}.}
\label{drv}
\end{figure}

For LkCa~15, we did not detect any planets either in the total-intensity or in the inverted polarimetric images. Detecting planets is challenging when the protoplanetary disk is very bright, as in the case of LkCa~15. Nonetheless, we attempted to search for planets that might be hidden in the disk by removing the disk signal. This involved creating a median-smoothed version of the disk using an 8-pixel box size (twice the typical planet PSF extent of $\sim$4 pixels, i.e. $\sim$0.048$\arcsec$), which we then subtracted from the original image together with the reference star PSF, as detailed in Section~\ref{s2}. Although we successfully suppressed a significant fraction of the disk emission, residuals remain, as evident in Figure~\ref{drv}. Despite this partial removal, no new planet was detected.

In spite of the visual non-detection of planets, we place upper mass limits on unseen companions by deriving a 5,$\sigma$ contrast curve from the disk-subtracted image (Figure~\ref{drv}). To create this image, we first applied a 4-pixel box median filter to our best-fit disk model, then reran the \texttt{LOCI} algorithm using a composite reference PSF made up of the stellar PSF plus the filtered disk. Finally, the 5,$\sigma$ contrast at each radius was computed as the standard deviation within an annulus, normalized by the peak stellar flux measured from 2-second exposures of LkCa~15. The contrast curve as a function of angular separation is shown in Figure~\ref{cc}. Using \texttt{DUSTY} models and assuming an age of $\sim$1~Myr for the system \citep{2021AJ....161..114S}, we estimate the upper mass limits for planets based on the contrast magnitude, as shown on the right-hand $y$-axis of Figure~\ref{cc}. We can detect planets more massive than 2.5~M${J}$ outside $\sim$20~au and 1.5~M${J}$ outside 50~au. The deepest contrast is achieved between $\sim$200 and 250~au, where the detection limit is $\sim$1.36~M${J}$. However, the disk around LkCa~15 is very bright and the detection limit across the brightest part of the disk is around 3.62~M${J}$, so planets may remain hidden in regions of strong disk emission. It is important to note the considerable uncertainty ($\sim$5~Myr) in age estimates for such young systems, which in turn affects the reliability of companion mass limits derived from evolutionary models.

\section{Discussion}
\label{s4}
Based on the evidence from the scattering phase function, we have found a reasonable models for our protoplanetary disk LkCa~15 both in the scattered light as well as in the millimeter regime. We find that the smaller micron-sized grains are mainly in the inner region of the disk, while the millimeter-sized grains mainly populate the outer part of the disk from 58 au extending till 127 au. Thus, the radial segregation of the grains is clearly evident for our proto-planetary disk which is quite unusual as larger grains are expected to lie closer to the central star \citep{2014ApJ...780..153B}. This radial segregation of grain sizes in the disk could be attributed to some key mechanisms, a) there is more thermal energy close to the star which hinders the formation of larger-sized grains \citep{1996Icar..124...62P,2008ARA&A..46...21B}, b) the segregation of small and large dust grains is significantly influenced by gas pressure gradients \citep{2012A&A...546L...1D}. These gradients arise from variations in the disk's gas density and temperature.  Larger grains, being less coupled to the gas, drift towards regions of higher pressure where they get trapped. This occurs because pressure-supported gas needs less centrifugal force and thus less rotational velocity. The slow gas is like a `headwind' causing the large grains to drift inwards\cite{1977MNRAS.180...57W}. However, they can encounter pressure maxima in the disk (regions of higher gas pressure) that act as barriers, trapping them and preventing further inward migration \citep{2012A&A...538A.114P}. This trapping can occur at locations such as a) Edges of gaps in the disk carved by planets \citep{2006A&A...459L..17P} b) Snowlines (e.g., the water snowline) where the gas density and temperature change sharply \citep{2020MNRAS.492..210V}. On the other hand, smaller grains, which are more tightly bound to the gas, continue to follow its motion without being significantly affected by these pressure variations. This differential behavior leads to the observed segregation, with larger grains accumulating in high-pressure areas, a key factor in planetesimal formation \citep{2010A&A...513A..79B,2012A&A...545A.134M,2022AJ....164...60S,2023AJ....166...91S,2024AJ....167..270S}. However, recent ALMA surveys of gas tracers in LkCa~15 challenge scenarios involving strong pressure bumps. \cite{2022A&A...663A..23L}, using $^{13}$CO and C$^{18}$O J=2-1 data, report that the gas surface density inside the $\sim45$-60 au dust gap drops by only a factor of $\sim$ 2 which is far less than expected for a deep planet-carved cavity. Complementary $^{12}$CO J=3-2 observations from the exo-ALMA survey \citep{2025ApJ...984L..16G} show a radially smooth gas distribution, even within the dust cavity, without evidence of a sharp depletion. These findings indicate that any pressure maxima in LkCa~15 are relatively modest and unlikely caused by a single, massive planet.
Instead, the observations favor multiple lower-mass planets creating gentle pressure variations, or alternative, non-planetary mechanisms that maintain gas continuity while enabling partial dust segregation.

In addition to reproducing the disk's morphological features, our analysis of the scattered-light phase function and polarization fraction provides crucial, independent constraints on the physical properties of the grains at the disk surface. By extracting the scattering phase function from our $K_s$-band images using \texttt{diskmap}, we find that neither pure Rayleigh scatterers (i.e., submicron grains) nor compact micron-sized grains can account for the observed shape and amplitude of the phase curves. Instead, comparison to theoretical models of porous aggregates with varying porosity and monomer number reveals that the disk's surface is best described by fluffy, micron-scale aggregate grains specifically, the CAHP$-$128$-$100nm model, which consists of 128 submicron monomers and $\sim$87\% porosity. This population successfully reproduces both the moderate forward-scattering enhancement and the broad, sub-Rayleigh polarization peak observed, further supporting a scenario where aggregate growth and porosity, not just size, are key in shaping the observable disk properties.

\begin{figure}[t]
\centering
\includegraphics[width=1 \columnwidth]{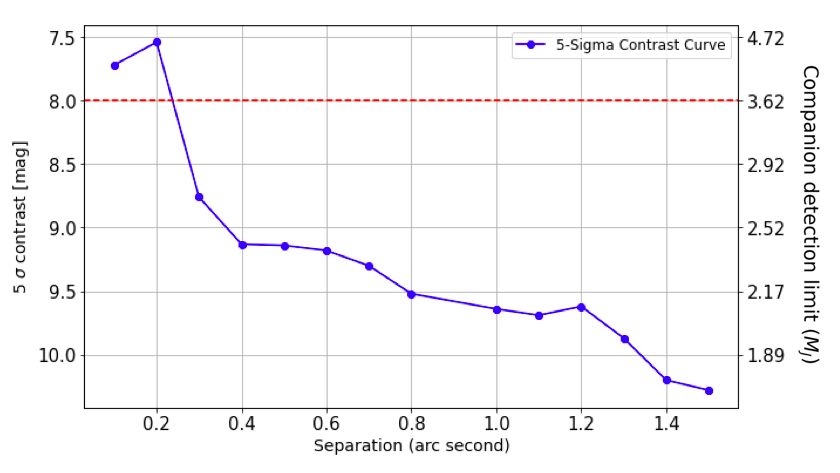}
\caption{The 5$\sigma$ contrast curve in $K$-band obtained by removing the LkCa~15 protoplanetary disk. The red dashed line represents the detection limit at the brightest part of the disks. The mass estimates for the companions were obtained using \texttt{DUSTY} models assuming the stellar age to be 1 Myrs for LkCa~15.}
\label{cc}
\end{figure} 

The classical collisional-cascade model of \citet{1969JGR....74.2531D}, originally developed for asteroidal and dust populations, predicts that the steady-state size distribution follows a power law
\begin{equation}
\label{gr}
    \frac{dN}{da} \propto a^{\zeta},
\end{equation}
where $dN/da$ is the number of particles per unit size interval, $a$ is the particle size, and $\zeta$ is the power-law exponent. In an equilibrium state where collisional fragmentation and coagulation balance each other, Dohnanyi found $\zeta = -3.5$. Although this model formally applies to gas-free collisional cascades, it has become a standard benchmark because $\zeta=-3.5$ successfully reproduces the grain-size distributions inferred for debris disks and the diffuse ISM \citep{1969JGR....74.2531D,1977ApJ...217..425M}, where the absence of gas makes the collisional physics comparatively simple and well constrained.

For our olivine model, with representative grain sizes of $12\,\mu$m and $2000\,\mu$m, the number ratio of grains predicted by Equation~\ref{gr} for the canonical slope $\zeta=-3.5$ is smaller by a factor of $\sim2\times10^{5}$ than the ratio implied by our best-fit model between these two populations. Treating these sizes as sampling a single power law, reproducing the measured ratio requires a much flatter slope of $\zeta\simeq -2.31$. In a gas-rich protoplanetary environment like LkCa~15, such a departure from the Dohnanyi value is not unexpected: gas drag, reduced collision velocities, efficient growth of larger aggregates, and radial drift can all act to deplete small grains and enhance the abundance of millimetre-sized particles \citep[e.g.][]{2010A&A...513A..79B,2014prpl.conf..339T}. Our inferred $\zeta\simeq -2.31$ is comparable to the shallower sub-micron-to-sub-millimetre slope ($\zeta\sim-2.7$) recently derived for PDS~70 \citep{2024A&A...687A.257W}, suggesting that flattened size distributions between micron and millimetre scales may be a common feature of gas-rich disks rather than an anomaly specific to LkCa~15.

\begin{figure}[t]
\centering
\includegraphics[width=1\columnwidth]{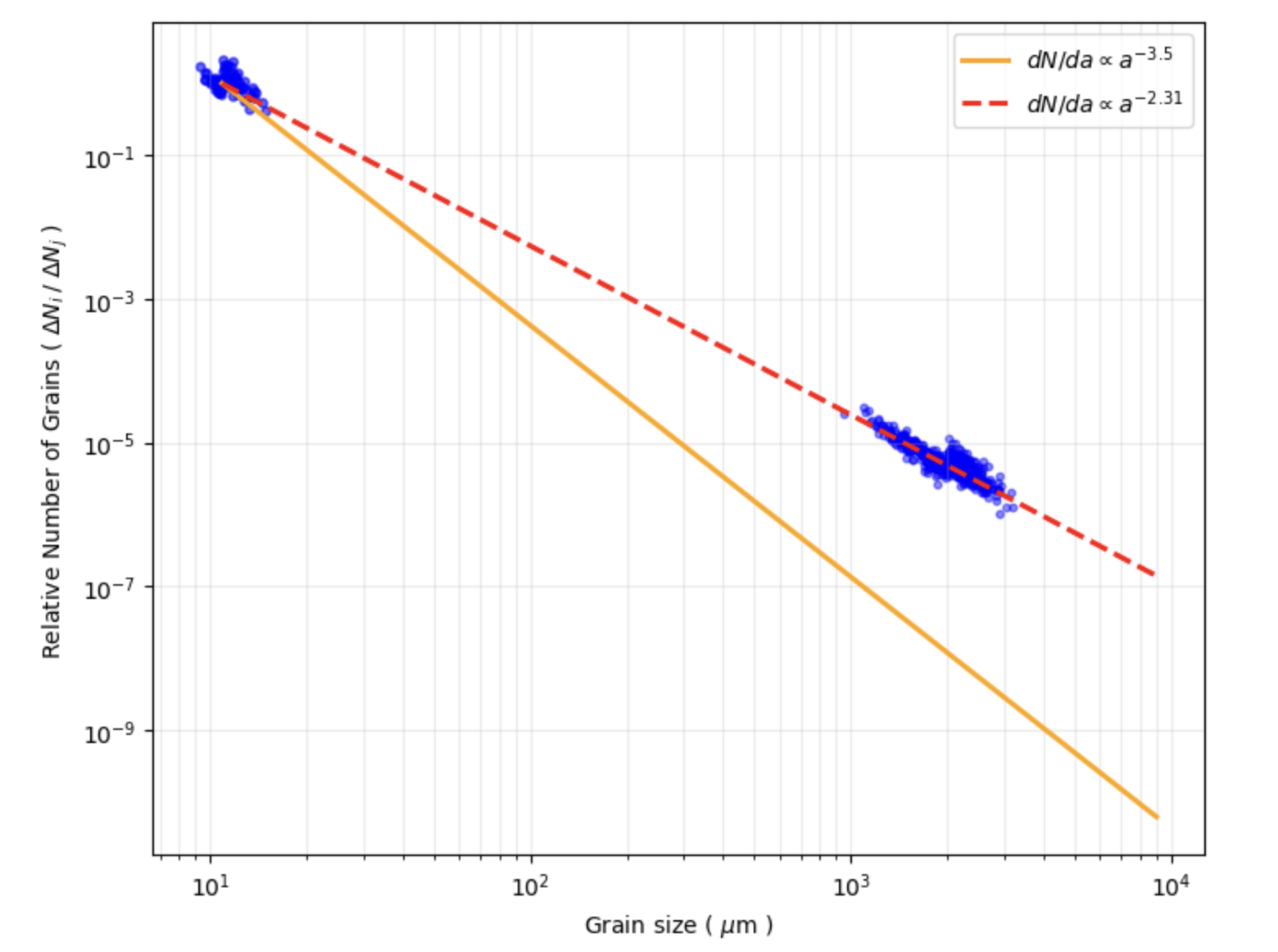}
\caption{Figure illustrates the model-estimated relative abundances of three grain sizes, as compared to those predicted by the power-law by \cite{1969JGR....74.2531D}. The data pairs are generated a Monte Carlo simulation with an asymmetric Gaussian distribution (see section~\ref{s4} for more details.)}
\label{gsd}
\end{figure}

Material properties may also contribute to this deviation. Classical Dohnanyi cascades assume asteroidal material composed of silicates, nickel, and iron, whereas our radiative-transfer models use amorphous olivine grains. Differences in composition and tensile strength can influence collisional outcomes and fragment size distributions. \citet{2024A&A...687A.257W} proposed several explanations for the apparent lack of small grains in PDS~70, including the migration of sub-micron grains into the disk gap and accelerated growth of millimetre grains in the midplane. The latter mechanism naturally increases the relative abundance of mm-scale grains and flattens the size distribution between microns and millimetres, qualitatively consistent with our inferred slope. In Figure~\ref{gsd}, the blue points represent grain sizes sampled from a distribution inferred from our model dust masses; grain-size and mass uncertainties are propagated via a Monte Carlo approach with an asymmetric Gaussian distribution, following Appendix~F of \citet{2024A&A...687A.257W}. Extending this kind of joint imaging–plus–size-distribution modelling to a larger sample of protoplanetary disks will be essential to determine whether shallow slopes like $\zeta\simeq -2.3$ are typical in gas-rich systems.

While we successfully obtained an artefact-free image of LkCa~15, previous claimed detections of proto-planets within this system remains unconfirmed \citep{2019ApJ...877L...3C}. This raises a key question: Are the planets hidden, or is the system too young or not massive enough for the formation of massive giant planets? The brightness of our disk also limits our detection capabilities in the disk-dominated regions as can be seen from the contrast curve in Figure~\ref{cc}. In comparison to systems like PDS 70, LkCa~15 is considerably younger, with an estimated age of approximately $1-2$ Myrs. Since both PDS~70 and LkCa~15 are pre-main sequence stars and have protoplanetary disks and are thus in a similar evolutionary stage, by assuming a mass accretion rate for LkCa~15 equal to that of PDS 70 \citep{2018ApJ...863L...8W}, a ranges from $10^{-8} M_{J}/yr$ to $10^{-7} M_{J}/yr$, we infer that the maximum mass of a forming planet in this system would be around $0.2 M_{J}$. Although this method of estimating planetary mass is somewhat oversimplified, it suggests that planets within LkCa~15 could be in the early stages of formation and may still be too small to be detected by high-contrast imaging techniques.


\section{Conclusions}
\label{s5}
In this work, we obtained $K_{s}$-band total-intensity and polarized-intensity images of the LkCa~15 protoplanetary disk using the star-hopping RDI technique. Our observations capture the intrinsic disk intensities for the first time, without the artefacts present in previous ADI-based observations. The primary goals of this work were (a) to search for new planets in the $\sim$20--30~au region using both total-intensity images and fractional-polarization maps, and (b) to understand the disk morphology and probable dust composition. We analysed the scattering phase function and performed radiative-transfer modelling with \texttt{RADMC-3D}. In particular, we attempted to reproduce the total- and polarized-intensity images as well as the ALMA millimetre continuum in order to obtain a complete picture of the disk. We were able to constrain most of the observed disk morphology. We summarise the key findings as follows:
\begin{itemize}
    \item We first modelled the disk with micron- and millimetre-sized olivine grains. This setup reproduces the total-intensity $K_{s}$-band image and the ALMA continuum reasonably well, but the scattered-light images are too strongly forward-scattering and too faint on the back side compared to the data.

    \item The observed intensity and polarization phase curves are inconsistent with pure Rayleigh scatterers or compact olivine grains. Instead, they are best matched by the \texttt{CAHP-128-100}\,nm porous-aggregate model from \texttt{AggScatVIR}, indicating that the disk surface is dominated by fluffy, micron-sized aggregates. 

    \item Using \texttt{CAHP-128-100}\,nm grains, guided by the scattering phase function, our model reproduces the key observed morphological features in scattered light, including the ``ear''-like lobes in the $Q$ and $U$ images and the observed level of forward scattering that olivine grains could not match. 

    \item Our best-fit model consists of a micron-grain surface layer from 25--85~au and a millimetre-grain midplane ring from 55--130~au.

    \item From the number ratio between the $\sim$12~$\mu$m and $\sim$2~mm grain populations, we infer an effective size-distribution slope of $\zeta \simeq -2.3$, implying that micron-sized grains are under-abundant compared to ISM/debris-disk-like distributions. 

    \item There may be radial drift in the dust grains, with small grains lying mainly within the inner 50~au of the disk, while the outer part of the disk is mainly composed of larger millimetre-sized grains. This radial segregation is unusual relative to standard expectations for grain growth and drift. 

    \item We analysed the inverted polarized-intensity map but did not find any discernible planetary signatures in the high-SNR regions of the disk. 

    \item We could not detect any planets in the total-intensity images either. Despite not detecting new planets within the disk, we estimate upper mass limits for potential planets: planets more massive than $1.4\,M_{J}$ are unlikely to exist beyond 200~au, whereas the inner brighter disk regions may conceal planets up to $\sim3.6\,M_{J}$. 

\end{itemize}

In conclusion, compact olivine grains provide a good overall joint fit to the near-IR and millimetre data, while porous \texttt{CAHP-128-100}\,nm aggregates are required to reproduce the near-IR scattering phase function and scattered-light morphology. Since CAHP opacities are not yet available at millimetre wavelengths, modelling the ALMA continuum with CAHP grains must await future opacity calculations. 

We acknowledge the non-uniqueness of these solutions, given the degenerate nature and interplay of various parameters such as flaring, surface-density distribution, and pressure scale height. Furthermore, it is difficult to determine the dust grain composition of protoplanetary disks from imaging observations alone. Therefore, future efforts should aim for spectroscopic observations, which can provide more direct constraints on grain composition.

Finally, while we successfully obtained self-subtraction-free images of LkCa~15, previous claimed detections of protoplanets in this system remain unconfirmed. Given the youth of the system and the strong disk brightness, any embedded planets may still be in an early growth phase and below current detection limits. 

It is also crucial to recognize the computational limitations in determining the optimal solution. The long model computation times pose a challenge, confining us to the discovery of the best solution around local minima rather than the true global best-fit solution. Moreover, accurate modelling and prediction of protoplanetary disk morphology are impaired by uncertainties in the underlying physical conditions. Despite these caveats, our simple radiative-transfer model accounts for the main features observed in LkCa~15. 

\begin{acknowledgements}
S.C. is funded by the European Union (ERC, UNVEIL, 101076613). Views and opinions expressed are however those of the author(s) only and do not necessarily reflect those of the European Union or the European Research Council. Neither the European Union nor the granting authority can be held responsible for them. S.C. acknowledges financial contribution from PRIN-MUR 2022YP5ACE. This project has received funding from the European Research Council (ERC) under the European Union's Horizon 2020 research and innovation programme (ERC, PROTOPLANETS, 101002188). This project received funding from the Department of Science and Technology, India as PhD fellowship
of C. Swastik. R.T. is supported by JSPS KAKENHI Grant Numbers JP25K07351.
\end{acknowledgements}

\begin{appendix}
\section{The star-hopping pipeline: more detections from the Ren et al.\ 2023 sample.}
\label{A4}
\begin{figure*}
\centering
\includegraphics[width=1 \textwidth]{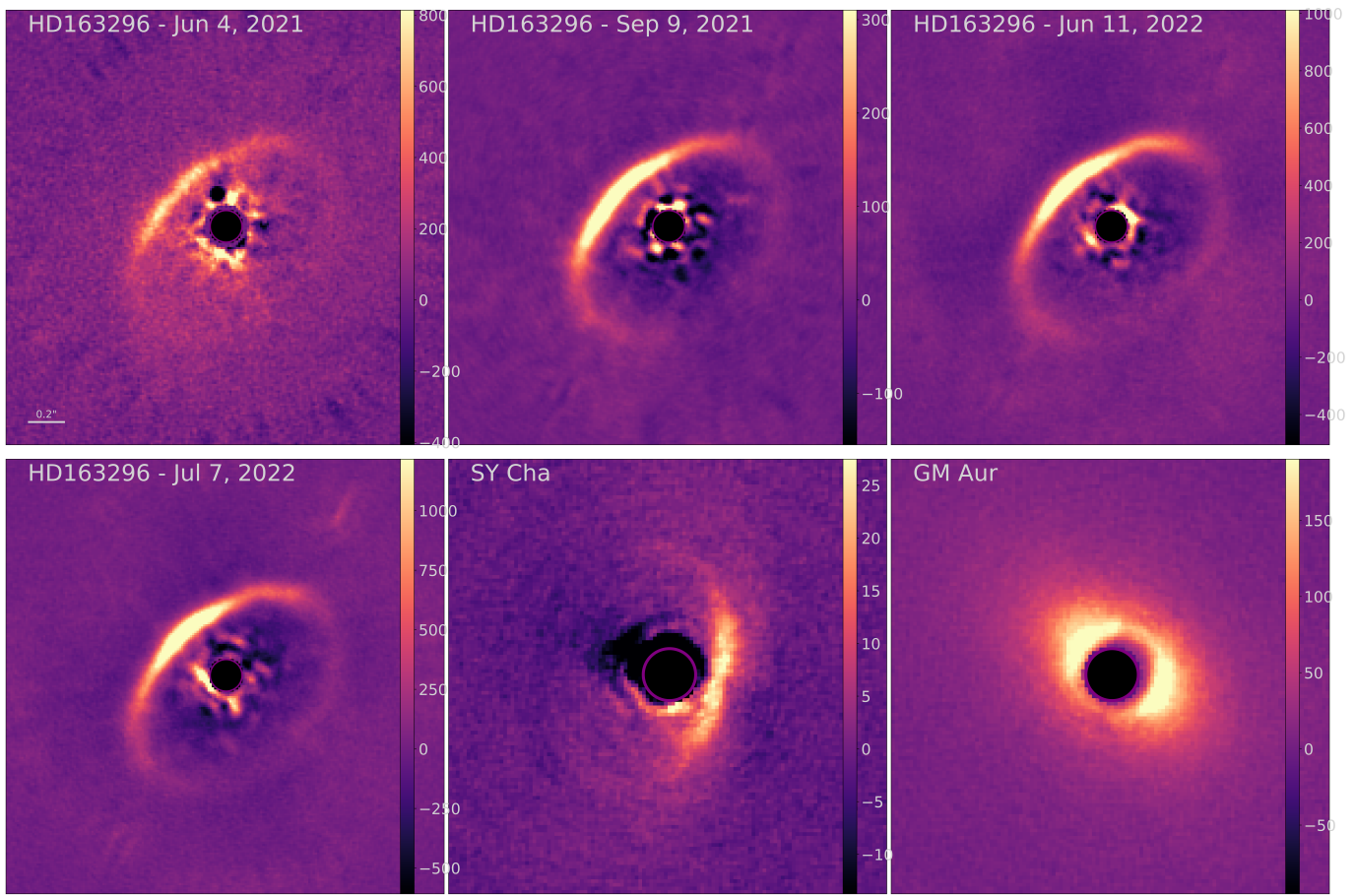}
\caption{Reduction of Ks-band protoplanetary disk using our star-hopping RDI pipleine. These fainter disks were not recovered by \cite{2023arXiv231008589R}.}
\label{rm3}
\end{figure*}

The star-hopping RDI reduction for disks is done using the steps below:  
\begin{itemize}
    \item {\bf Basic calibration:} Flat-field, remove bad pixels, and center the raw image files to obtain the science and reference image frames as done in  \cite{2021A&A...648A..26W}
    \item {\bf Reference radial profile correction:} Since the reference star might not have the same brightness as the science star, the reference  radial intensity profiles needs to be matched to the median science profile. We do this by multiplying a radial scaling function, \(\alpha(r)\), to the reference frames. The radial function, \(\alpha(r)\), is defined as:
    \[
    \alpha(r) = \frac{(c_1 - b_1)(r / r_0)^{-p} + b_1}{(c_2 - b_2)(r / r_0)^{-p} + b_2},
    \]
    where \(c_1\) and \(c_2\) are the fluxes of the science (median-combined images) and reference frames, respectively, at the coronagraphic edge (\(r = r_0\)), and \(b_1\) and \(b_2\) are the fluxes at the AO correction radius (\(r = r_1\)). The scaling ensures that \(\alpha(r_0) = c_1 / c_2\) and \(\alpha(r_1) = b_1 / b_2\), matching the fluxes at these key radii.

    The power-law index, \(p\), was determined from the radial profile of the median science frames using:

    \[
    c_1 \left(\frac{r_1}{r_0}\right)^{-p} = b_1,
    \]

    yielding:

    \[
    p = -\frac{\log(b_1 / c_1)}{\log(r_1 / r_0)}.
    \]

    The derived scaling function, \(\alpha(r)\), was applied to the reference frames to match the median radial profiles of the science frames, ensuring consistency at \(r_0\) and \(r_1\).

    \item \textbf{Remove background slopes:} Fit and remove an inclined plane from the science and reference images to mitigate suboptimal sky subtraction.
    \item \textbf{Identify signal and background regions:}
    \begin{itemize}
        \item \textbf{Preliminary science reduction:} Perform a LOCI PSF subtraction \citep{2007ApJ...660..770L} within the defined region using the most basic configuration-a full‐frame, single-region with no radial or azimuthal sub-sectors so that the PSF model is derived by minimizing residuals across the entire annulus in one step.
        \item -- \textbf{Reference-alone ADI:} Perform a classical ADI reduction of the reference images independently and use the residuals as a detection limit for the LOCI science reduction. Use this detection threshold to identify signal and background regions in the science reduction.
    \end{itemize}
    \item \textbf{Science cube median PSF subtraction:}
    \begin{itemize}
        \item \textbf{Rank reference PSFs:} Define a region between 8--70 pixels from the star, corresponding to the primary detection area. For each science image, perform optimally intensity-scaled subtractions with all reference images and identify those that yield an RMS within 30\% of the minimum RMS in the residuals. This provides an optimal reference set for each science image, used in subsequent steps.
        \item \textbf{Make science 2nd cube:} Find LOCI PSF matches for the median combination of the science frames on each side of the IRDIS detector, using the previously defined background region and the optimal references. Subtract the reference PSF match for each detector side from the corresponding science image to produce a difference cube, $s_1$.
        \item \textbf{Make reference 2nd cube:} Subtract the same reference PSF match from each reference image for its detector side to create a difference cube, $r_1$.
        \item \textbf{Individual science PSF subtraction:} Subtract LOCI PSF matches from each image in $s_1$ using a linear combination of images in $r_1$, restricted to the defined background region and using the optimal references, to create a difference cube, $s_2$.
    \end{itemize}
    \item \textbf{De-rotate and combine:}
    \begin{itemize}
        \item \textbf{De-rotation and SNR calculation:} De-rotate each image in $s_2$ and calculate the signal-to-noise ratio (SNR) of the disk as the ratio of the total flux in the signal region to the standard deviation in the background region.
        \item \textbf{Median combination:} Median combine the de-rotated $s_2$ images, using only those with at least 70\% of the maximum SNR.
    \end{itemize}
\end{itemize}

 Using the pipeline above, we were able to successfully detect four faint disks that were not reported in  \cite{2023arXiv231008589R} for their star-hopping observations. Thus, we present here the images of the protoplanetary disks of HD 163296 (across four epochs), SY CHA and GM AUR (see Figure~\ref{rm3}).  We also have a disk detection for V1094, but deem it spurious. This false detection is caused by the star moving behind the coronograph by many pixels, due the differential tip-tilt loop opening. Thus, we do not shows this reduction. We point out that if the reductions show strong azimuthally symmetric arcs in the background region, it is likely due do strong remaining speckles which are rotationally smeared. If there are large smooth positive or negative regions, this is likely due to incorrect intensity scaling of the reference PSFs with respect to the science PSFs. These disks were also detected in \cite{julliard_2024arXiv240614444J} by combining ADI and RDI techniques.

The technique in \cite{2023arXiv231008589R} separated circumstellar signals by flagging signal regions as “missing data” using a binary mask based on prior knowledge of their location. This ensures minimal contamination and avoids overfitting. In contrast, this work dynamically identified signal and background regions using a detection threshold obtained from a simple ADI reduction of the reference images only, requiring no prior knowledge of the signal location. 

\section{Estimation of random and systematic errors in disk recovery}
\label{A3}
To analyze the error maps and recovery fraction of the LkCa~15 protoplanetary disk, we used the HR8799 dataset obtained from \cite{2021A&A...648A..26W}. Since HR8799 lacks a protoplanetary disk component in the infrared (but a faint debris disk detected recently \citep{2024A&A...686A..33B}), we used this system as a test case to determine the recovery fraction of our starhopping pipeline (discussed in section~\ref{spp}). In order to calculate the recovery fraction, we introduced a simulated disk with different brightness into the HR8799 science dataset and subsequently recovered it using our pipeline. From there, we computed the recovery fraction as well as the systematic and random error maps for the datasets.

In simple terms, we made simulated disks of different contrasts from the rms map of the ``scaled reference" image. Here we discuss the detailed step-by-step explanation for generating the simulated disk and estimating the error maps:
\begin{enumerate}
    \item \textbf{Scaled Reference Images (sref): } Each sref image, generated during the LOCI reduction, serves as the best reference for corresponding science images.
    \item \textbf{Difference Image Map (dimg): } We subtract the median of the sref images from each individual sref image to create a difference image map. This is calculated as dimg$_{i}$ = sref$_i$ - median(sref).
    \item \textbf{Derotation: }To ensure sufficient derotation, we derotate each dimg$_{i}$ image by an angle of i × 0.1 degrees. We do not use the real derotation for the dataset, as these would recover the real planets and possibly other astrophysical signal.
    \item \textbf{RMS Map Construction: }The root mean square (RMS) map is constructed from the median combination of the dimg images. To be successfully detected, simulated signal has to be inserted above the noise level defined by this rms map.
    \item \textbf{Simulated Disk Creation: }Using the RMS map, we create a disk with an inner radius of 10 pixels and an outer radius of 80 pixels. This disk is then scaled in intensity and inserted into the science image to simulate proto-planetary disks.
    \item \textbf{Disk Recovery and Error Analysis: }We use the HR8799+simulated disk system as an input for our pipeline and recover the simulated disk. In order to evaluate the precision of disk recovery, we calculate fractional error maps using the following equation: \begin{equation} \text{Error map} = \frac{\text{Inserted disk} - \text{Recovered disk}}{\text{Inserted disk}} \end{equation}
    \item \textbf{Repetition and Scaling: }We repeat the entire process multiple times, scaling the RMS map by a set of multiplicative factors (n = 1, 3, 5, 7, 10, 20, 30, 40, 50, 75, and 100) to assess the impact of varying disk brightness on the recovery fraction.
\end{enumerate}

\begin{figure}[t!]
\centering
\includegraphics[width=1 \columnwidth]{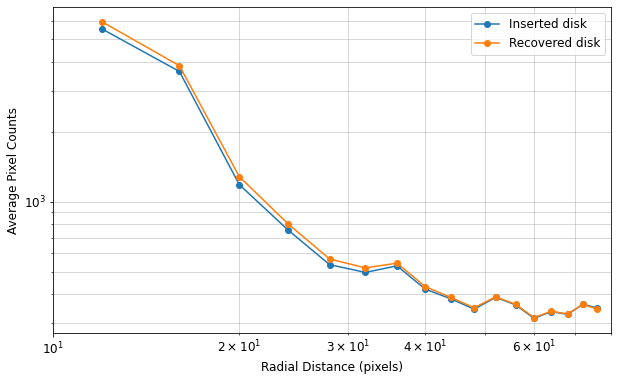}
\caption{Comparison of the radial profile of the simulated (blue) and recovered disk (orange) images at 30 $\sigma$ brightness level.}
\label{diskcomp}
\end{figure}

\begin{figure}[t!]
\centering
\includegraphics[width=1 \columnwidth]{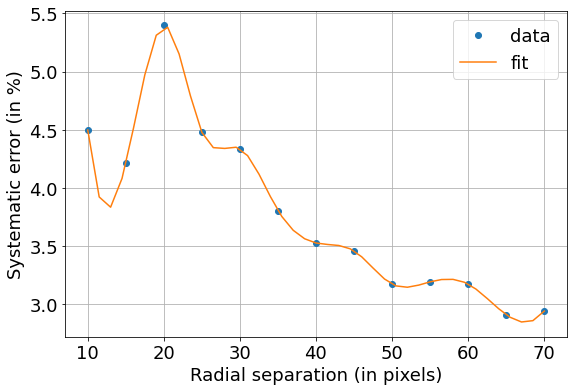}
\caption{Systematic error map for a disk of brightness 30 $\sigma$. }
\label{sm}
\end{figure}

\begin{figure}[t!]
\centering
\includegraphics[width=1 \columnwidth]{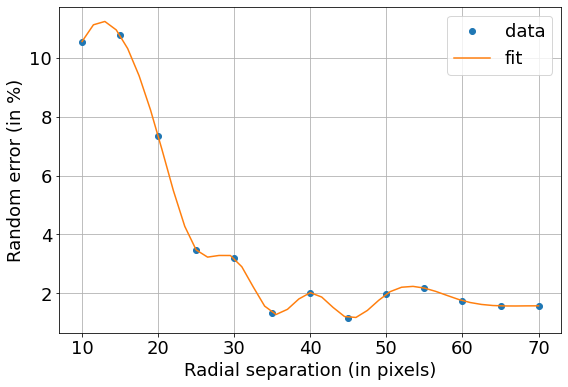}
\caption{Random error map for a disk of brightness 30 $\sigma$. }
\label{rm}
\end{figure}

Once we have the recovery maps for different contrast levels of the inserted disk, we use them to construct random and systematic error maps. The recovery map is divided into sub-annuli of 4 pixels, and the mean (which represents the systematic error) and standard deviation (which represents the random error) of pixel counts in each annulus are computed. Figure~\ref{diskcomp} shows the radial profiles of the simulated disk (inserted with 30$\sigma$ brightness) alongside the corresponding profile recovered using the star-hopping pipeline. We find that there is no significant flux loss or morphology change in the disk. We also observe that both the systematic and random error decrease with increasing radial separation as shown in Figures~\ref{sm} and ~\ref{rm}.

\begin{figure}[t!]
\centering
\includegraphics[width=1 \columnwidth]{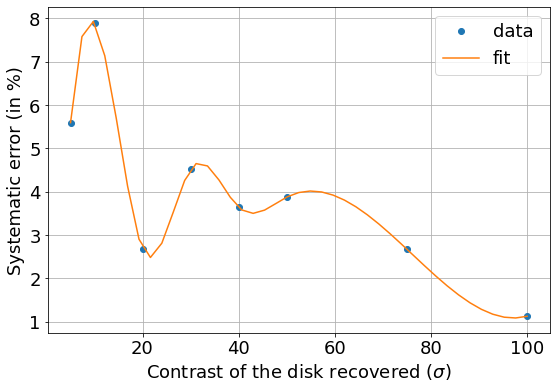}
\caption{Systematic error map as a function of disk brightness calculated in the annular region of 20 to 30 pixels from the central star.}
\label{rm2}
\end{figure}

\begin{figure}[t!]
\centering
\includegraphics[width=1 \columnwidth]{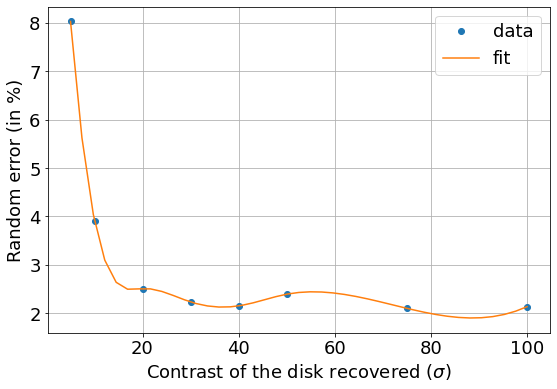}
\caption{Random error map as a function of disk brightness, calculated within the annular region spanning 20 to 30 pixels from the central star.}
\label{sm2}
\end{figure}

Furthermore, we find that both the random and systematic error decrease as the brightness of the disk increases. Figures ~\ref{rm2} and ~\ref{sm2} show this trend for a region between 20 and 30 pixels. Thus, for a high SNR dataset like that of LkCa~15 in this work ($\gtrsim
$ 150 $\sigma$), the error bars are on the order of $\leq$ 4$\%$ for the disk.

\section{Q$_{\phi}$ and U$_{\phi}$ images}
\label{A0}
Polarimetric pipelines like IRDAP produce azimuthal Stokes frames \(Q_{\phi}\) and \(U_{\phi}\) by rotating the \(Q/U\) maps such that \(Q_{\phi}\) captures polarization perpendicular (or parallel) to the radial direction and \(U_{\phi}\) captures the 45°‐offset component \citep{2020A&A...633A..63D}. While \(Q_{\phi}\) often exhibits higher signal to noise because it concentrates azimuthal disk signal into one frame, both \(Q_{\phi}\) and \(U_{\phi}\) together retain exactly the same information as the original \(Q/U\) pair (no Stokes information is lost by this basis rotation). However, for inclined and flared disks, multiple scattering and three dimensional surface geometry can introduce genuine non azimuthal polarization manifest as astrophysical signal in \(U_{\phi}\) and even negative \(Q_{\phi}\) values up to several percent of the total polarized flux (e.g., up to \(\sim4.5\%\) in generic \texttt{MCFOST} models at \(i=40^\circ\)) \citep{2015A&A...582L...7C}. These non‐azimuthal signatures are washed out if one examines \(Q_{\phi}\) alone, potentially biasing inferences of flaring or disk asymmetry.

\begin{figure}[t!]
\centering
\includegraphics[width=1 \columnwidth]{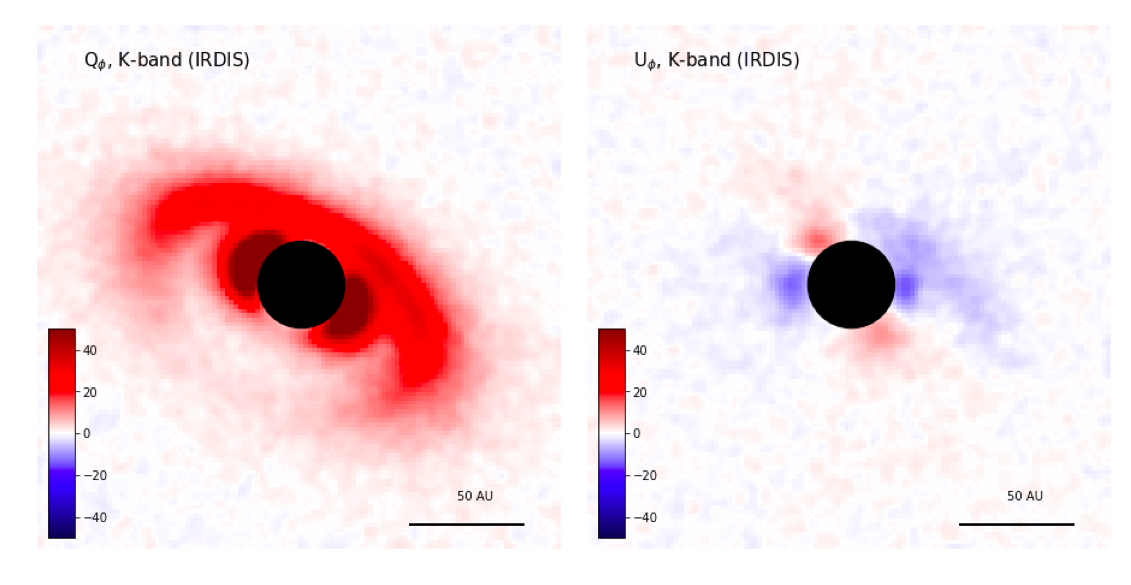}
\caption{The Q$_{\phi}$ and U$_{\phi}$ images are produced by the \texttt{IRDAP} pipeline, as described by \cite{2020ascl.soft04015V}.}
\label{UQphi}
\end{figure}

In contrast, the Stokes \(Q\) and \(U\) images preserve the full polarization vector at each pixel, allowing direct pixel by pixel comparison to RADMC‐3D's synthetic \(Q\) and \(U\) outputs without ambiguity from an assumed azimuthal pattern. Therefore, we chose to focus on modeling the Q and U images, as they more effectively represent the disk's geometry. For completeness, the Q${\phi}$ and U$_{\phi}$ images are shown in Figure~\ref{UQphi}.

\section{Calculations of $\overline{\chi}_{R}^2$}
\label{A1}
To obtain the best-fit solution, we need to compute the mean of the reduced chi-square ($\overline{\chi}_{R}^2$) for our images. The ($\overline{\chi}_{R}^2$) in turn consists of both the mean image reduced chi-square ($\overline{\chi}_{R,Img}^2$) and the mean flux chi-square ($\overline{\chi}^2_{f}$) which can also be expressed as,
\begin{equation}
\label{eqq}
   \overline{\chi}_{R}^2 =  \overline{\chi}_{R,Img}^2 + \overline{\chi}_{f}^2
\end{equation}
To calculate $\overline{\chi}_{R,Img}^2$, we first calculate the image chi-square ($\chi_{Img}^2$) in the region (rgn) from 10 to 75 pixels, considering the pixels with counts above 2 times the standard deviation in the background region (2$\sigma$). The $\sigma$ is determined by calculating the robust sigma in the background region between 75 to 100 pixels. In the above equation~\ref{eqq}, the model images are scaled before computing the $\chi_{Img}^2$. 
Thus the $\chi_{Img}^2$ is given as,
\begin{equation}
     \chi_{Img}^2 = \sum \frac{(data-model)^2}{\sigma^2} * \frac{nsig}{tp}
\end{equation}
where, the $nsig$ is the number of effective data-points in the $rgn$. The $nsig$ is obtained by dividing the total number of pixels in the $rgn$ with counts greater than 2$\sigma_{data}$ by the resolution area of each pixel and $tp$ is the total number of pixels in the $rgn$. In the above calculation, the model images are scaled before calculating $\chi_{Img}^2$. The scaling factor is determined by minimizing the standard deviation of the residuals between the observed and modeled data within a selected region (10 to 70 pixels). The scaled model is then computed by multiplying the initial model image by the scaling factor. Now, to get the $\chi_{R,Img}^2$, we need to divide $\chi_{Img}^2$ by $nsig$ for all our images as given by,
\begin{equation}
     \chi_{R,Img}^2 =  \frac{1}{nsig}*\chi_{Img}^2
\end{equation}

In order to measure the total flux mismatch, for each image the $\chi_{f}^2$ is calculated as,
\begin{equation}
    \chi_{f}^2 = \Bigg(\frac{ f_{data}-f_{model}}{A* f_{data}}\bigg)^2 
\end{equation}
where the uncertainties in science fluxes are taken to be around 15\% for NIR and 10\% for ALMA, $f_{data}$ and $f_{model}$ are the total data and model fluxes in the $rgn$. Since the polarimetric images contain both positive and negative fluxes, we needed to use the absolute values of the fluxes to calculate the $\chi_{f}^2$ for the Q and U images. Thus our final chi-square ($\overline{\chi}_{R}^2$) which we used for the minimization can be expanded as,
\begin{equation} \label{e3}
\begin{split}
\overline{\chi_{R}^2} = ( \chi_{R,I}^2+ \chi_{R,Q}^2+ \chi_{R,U}^2 +\chi_{R,880}^2 +\chi_{R,1300}^2    \\
+\chi_{f,I}^2+\chi_{f,Q}^2+\chi_{f,U}^2+\chi_{f,880}^2+\chi_{f,1300}^2)/10
\end{split}
\end{equation}
where the subscript $R$ and $f$ represent the reduced and flux chi-squares and the I, Q, U, 880, and 1300 represent the I, Q, and U images in the $K_{s}$-band, while 880 and 1300 represent the ALMA images in 880$\mu$m and 1300$\mu$m. 

To obtain the best-fit model, we begin with the initial parameters from the literature \citep{2019ApJ...881..108J,2021AJ....161..114S} and used the \textit{Basin-hopping} minimization algorithm from the Scipy.optimise\footnote{\url{https://docs.scipy.org/doc/scipy/reference/generated/scipy.optimize.minimize.html}} Python package, minimizing the $\overline{\chi_{R}^2}$ calculated in equation~\ref{e3} to derive the optimal solution for our data. Due to the computational demands of the minimization search process in \texttt{RADMC-3D}, it takes approximately 8 hours to complete approximately 500 iterations, making it impractical to find an exact best-fit solution for the data within these constraints. Instead, the solution obtained after 10000 iterations represents the closest approximation achievable. The $\overline{\chi}_{R}^2$ and other minimization parameter results are listed in Table~\ref{chitab}.

\begin{table}
\caption{\label{chitab} Summary of $\chi^2$ parameters and the flux differences (in $\sim\/\%$) for the olivine and CAHP grain models.}
\centering
\begin{tabular}{lc}
\hline\hline
Parameters &  Values \\
\hline 
\hspace{35pt} Olivine model  \\
\hline
Mean reduced $\chi^2$ & 1.31   \\
Mean reduced $\chi^2$ (scattered light) & 1.58   \\
Reduced $\chi^2$ (I) & 1.36 \\
Reduced $\chi^2$ (Q) & 1.45  \\
Reduced $\chi^2$ (U) & 1.93  \\
Reduced $\chi^2$ (880$\mu$m) & 0.60  \\
Reduced $\chi^2$ (1300$\mu$m) & 1.23  \\
Flux $\Delta$I & 14$\/\%$  \\
Flux $\Delta$Q &  34$\/\%$ \\
Flux $\Delta$U &  43$\/\%$ \\
Flux $\Delta$ 880$\mu$m & $4\/\%$ \\
Flux $\Delta$ 1300$\mu$m & $5\/\%$   \\
\hline
\hspace{35pt} \texttt{CAHP} grains model \\
\hline
Mean reduced $\chi^2$ (scattered light) & 1.35   \\
Reduced $\chi^2$ (I) & 1.06  \\
Reduced $\chi^2$ (Q) & 1.24  \\
Reduced $\chi^2$ (U) & 1.74  \\
Flux $\Delta$I & 11$\/\%$ \\
Flux $\Delta$Q & 8$\/\%$  \\
Flux $\Delta$U & 3$\/\%$ \\
\hline
\end{tabular}
\tablefoot{  $\Delta X = \frac{X_{\rm model} - X_{\rm obs}}{X_{\rm obs}}\times100\%$, for $X=I,Q,U,\,880\mu\mathrm{m},\,1300\mu\mathrm{m}$.
}
\end{table}

\end{appendix}
\bibliography{biblio}{}
\end{CJK*}
\end{document}